\documentclass[prb,twocolumn,showpacs,superscriptaddress]{revtex4-1}
\usepackage{epsfig}
\usepackage{amsmath,amssymb}
\usepackage{graphicx}
\usepackage{footnote}
\usepackage[english]{babel}

\usepackage{grffile}

\usepackage{color}

\usepackage{dsfont}

\newcommand{\ceq}[1]{Eq.~(\ref{#1})}
\newcommand{\cfg}[1]{Fig.~\ref{#1}}

\newcommand{\pdag}{{\phantom{\dagger}}}

\begin{document}

\title{Magneto-electric spectroscopy of Andreev bound states in Josephson quantum dots}

\author{Nils Wentzell}
\affiliation{Faculty of Physics, University of Vienna, Boltzmanngasse 5, 1090 Wien, Austria}
\affiliation{Institute for Solid State Physics, Vienna University of Technology, 1040 Vienna, Austria}
\affiliation{Institut f\"ur Theoretische Physik and Center for Quantum Science,
Universit\"at T\"ubingen, Auf der Morgenstelle 14, 72076 T\"ubingen, Germany}

\author{Serge Florens} 
\affiliation{Institut N\'{e}el, CNRS \& Universit\'{e} Grenoble Alpes, BP 166, F-38042 Grenoble, France}

\author{Tobias Meng}
\affiliation{Institut f\"ur Theoretische Physik, Technische Universit\"at Dresden, 01062 Dresden, Germany}

\author{Volker Meden}
\affiliation{Institut f\"ur Theorie der Statistischen Physik, RWTH Aachen University 
and JARA Fundamentals of Future Information Technology, 52056 Aachen, Germany}

\author{Sabine Andergassen}
\affiliation{Faculty of Physics, University of Vienna, Boltzmanngasse 5, 1090 Wien, Austria}
\affiliation{Institut f\"ur Theoretische Physik and Center for Quantum Science,
Universit\"at T\"ubingen, Auf der Morgenstelle 14, 72076 T\"ubingen, Germany}

\begin{abstract}
We theoretically investigate the behavior of Andreev levels in a
single-orbital interacting quantum dot in contact to superconducting leads, 
focusing on the effect of electrostatic gating and applied magnetic field,
as relevant for recent experimental spectroscopic studies.
In order to account reliably for spin-polarization effects in presence of
correlations, we extend here two simple and complementary approaches that 
are tailored to capture effective Andreev levels: the static functional renormalization 
group (fRG) and the self-consistent Andreev bound states (SCABS) theory.
We provide benchmarks against the exact large-gap solution as well as NRG calculations 
and find good quantitative agreement in the range of validity. 
The large flexibility of the implemented approaches then allows us to analyze a sizable
parameter space, allowing to get a deeper physical understanding into the
Zeeman field, electrostatic gate, and flux dependence of Andreev levels in interacting 
nanostructures.

\end{abstract}

\maketitle

\section{Introduction}
Andreev bound states (ABS) in quantum dots connected to superconducting
electrodes have been an subject of active research in recent years, both
theoretically~\cite{Glazman1989, Beenakker1992, Bauernschmitt1994, Ishizaka1995,
Yeyati1997, Alastalo1998, Rozhkov1999, Clerk2000, Choi2000, Rozhkov2000,
Avishai2001, Cuevas2001, Kusakabe2003, Zazunov2003, Vecino2003, Choi2004,
Siano2004, Oguri2004, Sellier2005, Bergeret2006, Lopez2007, Sadovskyy2007,
Benjamin2007, Pala2007, DellAnna2007, Skoeldberg2008, DellAnna2008,
Governale2008, Futterer2009, Zazunov2009, Zazunov2009a, Meng2009a, Zazunov2010,
Luitz2010, vZitko2010, Eldridge2010, Lee2010, Lim2011, Yamada2011,
MartinRodero2012, Sadovskyy2012, Droste2012, Brunetti2013, Baranski2013,
Futterer2013, Brunetti2013, Koga2013, Yokoyama2014, Rentrop2014, Zonda2015,
Kirsanskas2015} and experimentally~\cite{Baselmans1999, Kasumov1999,
Morpurgo1999, Buitelaar2002, Buitelaar2003, Kasumov2003, vanDam2006,
Cleuziou2006, Jarillo-Herrero2006, Jorgensen2006, Jorgensen2007, Buizert2007,
Sand-Jespersen2007, Grove-Rasmussen2007, Eichler2007, Eichler2009,
Jorgensen2009, Hofstetter2009, deFranceschi2010, Pillet2010, Herrmann2010,
Franke2011, Maurand2012, Ryu2013, Bauer2013, Pillet2013, Bretheau2013,
Bretheau2013a, Kim2013, Chang2013, Kumar2014, Abay2014, Schindele2014,
Lee2014}. The understanding of ABS formation is not only of great interest for
their potential use in quantum information devices, but also because they
constitute a testbed for microscopic theories of nanostructures.
Indeed, transport measurements in the normal state (obtained under the
application of a sufficiently strong magnetic field to suppress the
superconductivity in the leads) allow to extract in principle the basic
parameters governing the quantum dot (local Coulomb interaction $U$, tunneling
rate $\Gamma$, level position $\epsilon$). These in turn determine the
dispersion of the ABS in the superconducting state as a function of electrical
gating, the superconducting phase difference $\phi$, or with respect to a
moderate magnetic field $B$.  Several attempts for a precise description of
ABS in quantum dots have been recently made in this
direction~\cite{Pillet2010,deFranceschi2010,Bretheau2013}, but only qualitative
agreement could be obtained. In particular, microscopic calculations based on
the widely-used self-consistent Hartree-Fock approximation are not trustworthy
except for the case of weak Coulomb interaction or large applied magnetic
fields~\cite{MartinRodero2012,Zonda2015}.

Alternative theories to mean-field approaches offer a tradeoff between simplicity 
and accuracy. The simplest techniques are based on static renormalization group ideas, 
and have been formulated both within a perturbative expansion in the effective Coulomb 
interaction in the framework of the functional renormalization group 
(fRG)~\cite{Karrasch2008,Karrasch2009,Karrasch2011,Luitz2012,Rentrop2014}, 
or around the large gap limit by a self-consistent Andreev bound state picture
(SCABS)~\cite{Meng2009a,Meng2009,Maurand2012}. 
Both techniques achieve surprisingly good agreement (in their range of validity) with 
full-scale numerical renormalization group (NRG) computations~\cite{Satori1992, Sakai1993, 
Yoshioka2000, Bauer2007, Tanaka2007, Bulla2008, Hecht2008, Tanaka2008, deSousa2009, 
Oguri2013, vZitko2015}, while their low numerical cost allows to efficiently explore 
the effective Andreev levels over the whole parameter space. While previous
analytical renormalization group calculations have mainly focused on the particle-hole 
symmetric case (i.e. at the center of the odd charge Coulomb blockade diamond)
and for zero magnetic field, we aim here at extending both the fRG and SCABS techniques 
to account for the full electric and magnetic tuning available in quantum dot devices.  
We will not consider here full second-order perturbation theory in the Coulomb
repulsion $U$. Although this technique provides excellent results at
particle-hole symmetry and zero magnetic field, once self-consistency on the 
effective pairing amplitude is properly taken into account~\cite{Meng2009,Zonda2015}, 
its accuracy is expected to degrade away from these two limits (in addition, a
proliferation of diagrams makes the technique more cumbersome to use in absence
of any symmetry).

The paper is organized as follows. In Sec.~\ref{sec:model} we introduce the basic
model of superconducting quantum dots, and describe how to obtain the position and weights 
of ABS from Green's function techniques in presence of a Coulomb repulsion. The 
model is then solved mathematically in the special limit of infinite gap in presence 
of both an external gate voltage and an applied magnetic field, which allows for 
a qualitative discussion of the physics.
In Sec.~\ref{sec:methods}, we 
briefly review the static functional renormalization group and the 
self-consistent Andreev bound state theory extensions to the case of
finite magnetic field. Finally, we discuss our results in Sec.~\ref{sec:results}, 
starting with the case of zero magnetic field before considering the complete
magneto-electric spectroscopy of the Andreev levels. The various methods are
tested against previous NRG results~\cite{Bauer2007}, 
in order to assess their validity range and possible breakdowns.

\section{Superconducting quantum dot model}
\label{sec:model}

\subsection{The superconducting Anderson Hamiltonian}

\begin{figure}
\begin{center}
\includegraphics[width=0.35\textwidth]{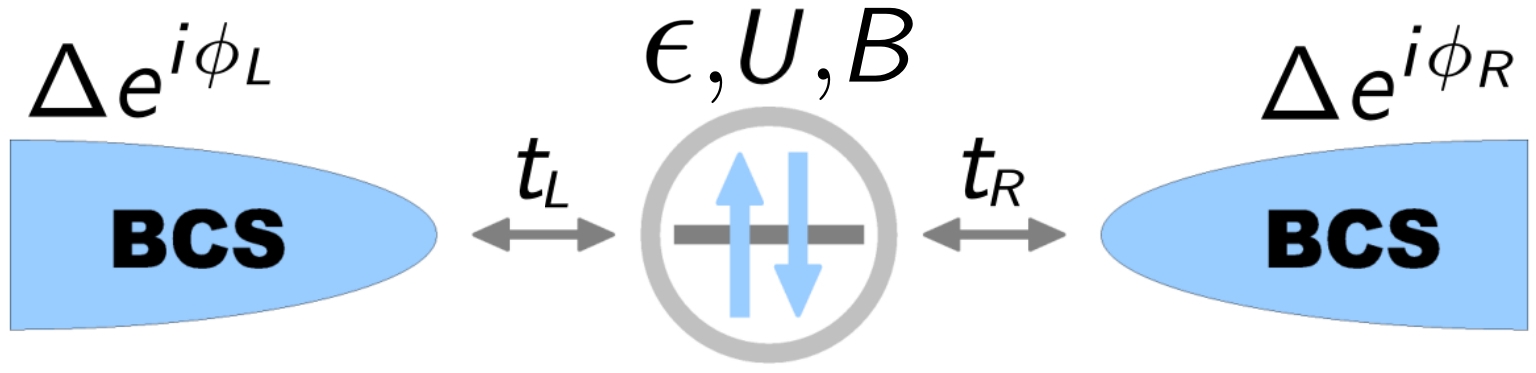}
\caption{Setup considered in this work. A quantum dot subject to a 
magnetic field ${B}$ and an electrical gate $\epsilon$ is
tunnel coupled to two superconducting BCS electrodes.} 
\label{fig:setup}
\end{center}
\end{figure}

Due to strong electronic confinement in quantum dots, it is legitimate to
base our study on a single-orbital level (exceptions arise however in
ultraclean carbon nanotube systems, where chirality and spin-orbit physics
can play an important role).
We assume here for simplicity that the magnetic field has no orbital effect
on the quantum dot (this applies for instance to the case of carbon nanotubes
that are perpendicular to the field axis) and only lifts the degeneracy between 
spin up and spin down states through the Zeeman effect. 
In the metallic leads, the Zeeman effect is usually negligible, but a sufficiently strong
orbital effect can suppress the superconducting gap. We will thus consider
here relatively weak magnetic fields, such that the superconducting order parameter 
(gap amplitude) $\Delta$ can be assumed 
constant. The possibility to tune the superconducting phase difference via the
magnetic field in a SQUID geometry will be accounted for via the independent
phase difference $\phi$ across the junction.  We thus investigate the model
depicted in Fig.~\ref{fig:setup}, that is described by the Hamiltonian
\begin{align}
\label{eq:Hamiltonian_complete_normal}
H &= \sum_{\alpha = L,R}{H_\alpha} + H^{\rm dot} + \sum_{\alpha = L,R}{H_{\alpha}^{\rm T}},
\end{align}
where
\begin{subequations}
\begin{align}
H_{\alpha}^{\rm} &= \sum_{\vec{k},\sigma}{\epsilon_{\vec{k}}^{} \, 
c_{\vec{k},\sigma,\alpha}^{\dagger} c_{\vec{k},\sigma,\alpha}^\pdag} 
- \sum_{\vec{k}}{\left( \Delta_{\alpha}^{} \, c_{\vec{k},\uparrow,\alpha}^{\dagger} 
c_{-\vec{k},\downarrow,\alpha}^{\dagger} + \mathrm{h.c.}\right) },
\label{eq:H_lead} \\
H^{\rm dot} &= \sum_\sigma\left(\epsilon \, d_{\sigma}^{\dagger} d_{\sigma}^{\pdag} 
+ \sigma B \, d_{\sigma}^{\dagger} d_{\sigma}^{\pdag}\right)
+ U \left(n_{\uparrow}-\frac{1}{2}\right) \left(n_{\downarrow}-\frac{1}{2}\right) \label{eq:H_dot}, \\
H^{\rm T}_{\alpha} &= \sum_{\vec{k},\sigma}{\left( t_{\alpha}^{} \, 
d_{\sigma}^{\dagger} c_{\vec{k},\sigma,\alpha}^{\pdag} 
+ \mathrm{h.c.}\right) } \label{eq:H_T_LD} .
\end{align}
\end{subequations}
In the above equations $\alpha=L,R$ denotes the left and right lead
respectively, while $\sigma=\uparrow,\downarrow$ denotes the spin degree of
freedom.  The leads are modeled by BCS Hamiltonians $H_\alpha$ with a
lead-independent dispersion $\epsilon_{\vec{k}}^{}$
and superconducting gaps $\Delta_\alpha = |\Delta|\,e^{i\phi_\alpha}$ that differ 
only in the complex phase $\phi_\alpha$. 
Note that only the phase difference $\phi =\phi_L-\phi_R$ is of physical importance.
We furthermore assume the leads to have a flat density of states of
amplitude $\rho_0 = 1/(2D)$, where $2D$ is the total bandwidth. 
The leads are tunnel coupled to the quantum dot by tunneling amplitudes $t_{
\alpha }$, which we assume to be momentum independent.  The dot, finally, is
characterized by a level energy $\epsilon$, an on-site Coulomb repulsion $U$,
and a Zeeman energy $B$. Note that the single-particle energy was shifted, 
such that $\epsilon=0$ corresponds to the particle-hole symmetric case.
As discussed above, the lead parameters (such as the superconducting gap $\Delta$ and the phase 
difference $\phi$) are considered to be effective parameters for a given 
magnetic field.

\subsection{Green's functions in superconducting dots}
For practical reasons, we will work in the following with the Nambu operator basis 
\begin{equation}
\Psi = \left( \begin{array}{c}d_\uparrow \\ d_\downarrow^\dagger \end{array} \right)
\end{equation}
for the dot degrees of freedom.
This allows us to introduce a matrix structure for all one-particle correlation
functions (defined below on the Matsubara imaginary axis), such that the off-diagonal 
terms capture the anomalous components, while the diagonal terms can be directly related 
to the normal spin-resolved ones:
\begin{align}
G(i\omega) 
= \begin{pmatrix} G_{11}(i\omega) & G_{12}(i\omega)\\
G_{21}(i\omega) & G_{22}(i\omega)
\end{pmatrix}
= \begin{pmatrix} \langle d_\uparrow d_\uparrow^\dagger \rangle_{ i\omega } 
& \langle d_\uparrow d_\downarrow \rangle_{ i\omega }\\
\langle d_\downarrow^\dagger d_\uparrow^\dagger \rangle_{ i\omega } 
& \langle d_\downarrow^\dagger d_\downarrow \rangle_{ i\omega }
\end{pmatrix}.
\end{align}
We first consider the situation of a non-interacting quantum dot ($U=0$). 
In the wide band limit, i.e. $D\rightarrow\infty$ while keeping the ratio 
$D/t^2$ constant, the Green's function of the dot level is given by
\begin{align}
\label{eq:Green}
G_0(i\omega) &= \begin{pmatrix} i\tilde{\omega} - \epsilon - B& \tilde{\Delta}
\\ \tilde{\Delta}^* & i\tilde{\omega} + \epsilon - B\end{pmatrix}^{-1} \nonumber\\
  &= \frac{1}{D_0(i\omega)} \begin{pmatrix} i\tilde{\omega} + \epsilon -B &
-\tilde{\Delta} \\ -\tilde{\Delta}^* & i\tilde{\omega} 
  - \epsilon - B\end{pmatrix}, 
\end{align}
with the determinant 
\[
D_0(i\omega)=\left( i\tilde{\omega} - \epsilon - B \right) \left(i\tilde{\omega} 
+ \epsilon - B \right) - |\tilde{\Delta} |^2.
\]
We also introduced the compact notations
\begin{align}
i\tilde{\omega} &= i\omega\left(1+\frac{\Gamma}{\sqrt{\omega^2 + \Delta^2}}
\right ), \\
\tilde{\Delta} & = \frac{\Delta}{\sqrt{\omega^2+\Delta^2}} \sum_{\alpha=L,R}
\Gamma_\alpha e^{i\phi_\alpha},%\\
\end{align}
with a total hybridization $\Gamma = \sum_{\alpha = L,R} \Gamma_{\alpha}$, and 
$\Gamma_\alpha = \pi \rho_{0} t_\alpha^2$. Note that $\Gamma$ will in the following 
be used as our unit of energy.

At the one-particle level, the effects of the local Coulomb interaction $U$ can 
be fully accounted for by a frequency-dependent self-energy, so that the interacting 
Green's function of the dot reads
\begin{equation}
\begin{split}
&G(i\omega) 
= \left(G_0^{-1}(i\omega) - \Sigma(i\omega)\right)^{-1} \\ 
&  = \frac{1}{D(i\omega)} \begin{pmatrix} i\tilde{\omega} + \epsilon - B
-\Sigma_{2}(i\omega) & -\tilde{\Delta} +\Sigma_{\Delta}(i\omega)
\\ 
  -\tilde{\Delta}^* + {\Sigma_{\Delta}}^*(-i\omega) & i\tilde{\omega}
- \epsilon - B - \Sigma_{1}(i\omega) \end{pmatrix},
\end{split}
\end{equation}
with the determinant
\begin{align}
\nonumber
  D(i\omega) =& \left[ i\tilde{\omega} - \epsilon - B -
\Sigma_{1}(i\omega) \right] \left[i\tilde{\omega} 
  + \epsilon - B - \Sigma_{2}(i\omega)\right]\\
& - |\tilde{\Delta} - \Sigma_{\Delta}(i\omega)|^2.
\end{align}

\subsection{Andreev bound states, spectral weights, and Josephson current}
\label{sec:ABS}

The density of states of the quantum dot features discrete ABS inside
the superconducting gap. They correspond to poles in the total electronic density of states
\begin{align}
\rho(\omega) =& -\frac{1}{\pi}\lim_{\eta\rightarrow 0^{+}} \Im \mathrm{m}\left[ G_{11}(\omega+i\eta) -
G_{22}(-\omega-i\eta) \right]
\label{DOS}
\end{align}
that can be determined by finding all roots $E_{\rm bs}\in\{\pm a, \pm b\}$ of the 
determinant $D(\omega)$ on the real frequency axis. Note that ABS poles will always appear 
in pairs symmetrically positioned around the chemical potential, while their respective 
spectral weights are calculated from their residuals
\begin{align}
w(E_{\rm bs})= \lim_{\eta\rightarrow 0^{+}}i\eta \, 
\left[ G_{11}(E_{\rm bs}+i\eta) - G_{22}(-E_{\rm bs}-i\eta) \right].
\label{NormWeight}
\end{align}
In addition we will consider the weight of the anomalous component of the
Nambu Green's function
\begin{align}
w_{\Delta}(E_{\rm bs})= \lim_{\eta\rightarrow 0^{+}}i\eta \, G_{21}(E_{\rm
bs}+i\eta),
\label{AnWeight}
\end{align}
which contains information on the supercurrent carried by the ABS.
As we will see in the following, the ABS are responsible for a substantial part
of the total Josephson 
current~\cite{Josephson1962, Bloch1970} that can flow through the device in the presence of a finite 
superconducting phase difference $\phi$. To illustrate this, let us define the Josephson current operator
as the time derivative of the particle number operator $N_\alpha$ for the left and right lead respectively
\begin{equation}
J_\alpha = \partial_t N_\alpha = i[H,N_\alpha].
\end{equation}
In the absence an applied bias and at $T=0$, the expectation value reads
\begin{equation}
\langle J_\alpha \rangle =\frac{2\Gamma_\alpha}{\pi} \int d\omega \;\Im\mathrm{m}\left[\frac{\Delta
e^{i\phi_\alpha}}{\sqrt{\omega^2 + \Delta^2 }} G_{21}(i\omega)\right].
\label{JC}
\end{equation}
This formula is valid also in presence of interaction, provided the exact
anomalous Green's function is known.
To determine the contribution of the different ABS to the
current, we split the Green's function $G_{21}$ into a part containing the poles,
and another part carrying the contribution of the spectrum corresponding to
branch cuts in the complex plane, which is associated to the continuum
above the gap:
\begin{equation}
G_{21}(i\omega) = G^{cont.}_{21}(i\omega) + \!\!\sum_{\{\pm E_{\rm bs}\}}
\frac{w(E_{\rm bs})}{i\omega - E_{\rm bs}}.
\end{equation}
Plugging this into Eq.~\eqref{JC} we obtain %($J = J_L$)
\begin{equation}
\langle J_L \rangle = \sum_{\{\pm E_{\rm bs}\}} \langle J_{E_{\rm bs}} \rangle + \langle J_{cont.}
\rangle,
\label{JCsplit}
\end{equation}
with
\begin{equation}
\label{integralJ}
\langle J_{E_{\rm bs}} \rangle = \frac{2\Gamma_L}{\pi} \! \int \! d\omega\;
\Im\mathrm{m}\left[\frac{\Delta e^{i\phi/2}}{\sqrt{\omega^2 + \Delta^2 }}
\frac{w(E_{\rm bs})}{i\omega - E_{\rm bs}}\right],
\end{equation}
and
\begin{equation}
\langle J_{cont.} \rangle = \frac{2\Gamma_L}{\pi} \!\int\! d\omega\;
\Im\mathrm{m}\left[\frac{\Delta e^{i\phi/2}}{\sqrt{\omega^2 + \Delta^2 }}
G_{21}^{cont.}(i\omega)\right].
\end{equation}
Evaluating the integral~(\ref{integralJ}) gives
\begin{equation}
\langle J_{E_{\rm bs}} \rangle = -2\Gamma_L \hspace{0.1cm} f\left(\left
|\frac{E_{\rm bs}}{\Delta}\right |\right) \mbox{sgn}\left( E_{\rm bs} \right )
\Im\mathrm{m} \left [e^{i\phi/2} w_\Delta(E_{\rm bs}) \right ],
\label{JCBS}
\end{equation}
where $f(x) = [\pi  - 2 \arcsin(x)]/(\pi\sqrt{1-x^2})$.  Note that the explicit
dependence of $\langle J_{E_{\rm bs}} \rangle$ on the relative bound state
position $\left |E_{\rm bs} / \Delta\right |$ is weak, so that the current
amplitude is mainly determined by the sign and weight of the ABS.

\subsection{The large gap limit}
\label{sec:LargeGap}

A simple physical picture of the ABS can be obtained from the limit \footnote{In the limit of infinite superconducting gap a local moment cannot be screened due to the lack of electronic states in the leads. Consequently, the Kondo effect is fully suppressed.} $\Delta \rightarrow \infty$. In this case, the non-interacting Green's function simplifies as
\begin{align}
\label{NonIntG}
G_0(i\omega)^{-1} 
\xrightarrow{\Delta\rightarrow\infty}
i\omega - \begin{pmatrix}   B + \epsilon  & -\Gamma_{\phi} \\  -\Gamma_{\phi}^*  &  B -
\epsilon  \end{pmatrix},
\end{align}
where $\Gamma_{\phi} = \sum_\alpha \Gamma_\alpha e^{i\phi_\alpha}$, which, for
the case of a symmetric coupling to the leads $\Gamma_L=\Gamma_R=\Gamma/2$,
takes the simple form
\begin{equation}
\Gamma_{\phi} = \Gamma_{\phi}^*  = \Gamma \;\rm{cos} \;\frac{\phi}{2}.
\end{equation}
The key point is that the non-interacting Green function~(\ref{NonIntG})
coincides with the one of a system with an effective local Hamiltonian
\begin{equation}
H_{\rm{\rm eff}}^0 =  \Psi^\dagger \begin{pmatrix} B + \epsilon & -\Gamma_{\phi} \\
-\Gamma_{\phi}^* & B - \epsilon \end{pmatrix} \Psi,
\label{ALHamNI}
\end{equation}
where $\Psi$ is the previously introduced Nambu spinor.
This Hamiltonian can be diagonalized by means of a Bogoliubov basis transformation
\begin{align}  
\label{nambu}
\Psi'=
\begin{pmatrix} d_+  \\ d^\dagger_- \end{pmatrix}
= 
\begin{pmatrix} u & -v \\ v^* & u^* \end{pmatrix}
\Psi,~
\end{align}
where $u$ and $v$ are defined up to an arbitrary phase factor by
\begin{subequations}
\begin{align}
u^*v&=\Gamma_{\phi}/(2E_{\phi}), \\
|u|^2&=(1+\epsilon/E_{\phi})/2, \\
|v|^2&=(1-\epsilon/E_{\phi})/2,
\end{align}
\end{subequations}
and 
\begin{equation}
E_{\phi} = \sqrt{\epsilon^2  + |\Gamma_{\phi}|^2}.
\end{equation}

The possibility to reduce the problem to a local one allows to deal with
the Coulomb interaction in a simple way.
In the new basis $\{|00\rangle, |01\rangle, |10\rangle, |11\rangle\}$, labeled by $(n_+,n_-)$, the 
full effective Hamiltonian takes
the diagonal form
\begin{align}
H_{\rm{\rm eff}}
= E_{\phi}(n_+ - n_-) + B (n_+ + n_- -1) + \frac{U}{2} (n_+ - n_-)^2,
\end{align}
with the eigenvalues
\begin{subequations}
\begin{align}
&E_{00} = -B, \quad &E_{01} &= E_{\phi} + \frac{U}{2},\\
&E_{10} = -E_{\phi}+\frac{U}{2}, \quad &E_{11}& = B.
\label{ATLIEigen}
\end{align}
\end{subequations}
The relations to the electronic dot-basis are shown in Table \ref{basis}. Here we 
have introduced the shorthands 
\begin{equation}
E_\sigma = \sigma B, \qquad E_{\pm} = U/2 \pm E_{\phi}.
\end{equation}
\begin{table}[b]
\centering
\begin{tabular}{c|c|l}
Eigenvalue & Eigenbasis &  Dot-basis \\
\hline $E_\uparrow$ & $|11\rangle$   & $\phantom{u^*}|\uparrow\rangle$ \\
\hline $E_\downarrow$ & $|00\rangle$ & $\phantom{u^*}|\downarrow\rangle$ \\
\hline $E_+$ & $|01\rangle$ & $|+\rangle = u^{\phantom{*}}|0\rangle + v^{\phantom{*}}|\uparrow\downarrow\rangle$ \\
\hline $E_-$ & $|10\rangle$ & $|-\rangle = v^*|0\rangle - u^*|\uparrow\downarrow\rangle$ 
\end{tabular}
\caption{Relations of the electronic dot-basis to the eigenbasis of the
effective interacting Hamiltonian.}
\label{basis}
\end{table}
Clearly (for positive $B$ and $U$, which we assume from now on), the system can 
assume only two possible ground states, either the non-magnetic $0$-phase state 
$|10\rangle$, or the spin polarized $\pi$-phase state $|00\rangle$.
A phase transition (level crossing) will occur %between both phases according to the condition
under the condition $E_\downarrow = E_-$, which reads explicitly
\begin{equation}
(U+2B)^2 =\; 4 \left[(\Gamma_R-\Gamma_L)^2+ 4\Gamma_L\Gamma_R \cos^2\frac{\phi}{2}\right]\nonumber + 4\epsilon^2.
\label{PTCond}
\end{equation}
This indicates the similar role of $U$ and $B$ in determining the phase
boundary, that is an increase of either parameters will induce a transition 
to the $\pi$-phase. However, an increase of $U$ alone will tend in addition
to renormalize strongly the electronic states on a wide energy range.

Using Lehmann's representation, %formula and the exact eigenstates, 
one can reconstruct the exact Green's function in the large gap limit (see 
App.~\ref{app:LargeGap}), and hence the corresponding
self-energies for finite magnetic fields $B \neq 0$
\begin{equation}
\Sigma = \begin{cases} \frac{U}{2E_{\phi}} \begin{pmatrix}
-\epsilon & \Gamma_{\phi} \\ \Gamma_{\phi}^* & \epsilon \end{pmatrix} & \quad0\mbox{-phase} \\
\qquad
\begin{pmatrix} \frac{U}{2} & 0 \\ 0 & \frac{U}{2} \end{pmatrix} &
\quad\pi\mbox{-phase} \end{cases}~~.
\label{SigFinB}
\end{equation}
Note that, in this exactly solvable limit, the self-energy is found to be frequency independent,
which is a strong argument for approaches that make the assumption of a static self-energy.
On the other hand, the self-energy is completely independent from the magnitude of the magnetic field, 
while being purely linear in $U$ in both phases. For finite magnetic field,
this is a strong argument in favor of approaches that are perturbative in $U$
(such as the static fRG or Hartree-Fock theory).

The situation changes drastically when we consider the case of vanishing magnetic field.
While the self-energy in the $0$-phase remains unchanged, the two-fold degeneracy of the ground state 
in the $\pi$-phase results in a frequency dependence as well as a $U^2$ scaling. At $B=0$ we find
\begin{equation}
\Sigma = \begin{cases} \qquad\quad \frac{U}{2E_{\phi}}
\begin{pmatrix} -\epsilon & \Gamma_{\phi} \\ \Gamma_{\phi}^* & \epsilon \end{pmatrix} &
\quad\!\!\!0\mbox{-phase} \\
\frac{U^2}{4} \frac{1}{(i\omega)^2 - E_{\phi}^2}  \begin{pmatrix} i\omega + \epsilon &
-\Gamma_{\phi} \\ -\Gamma_{\phi}^* & i\omega - \epsilon \end{pmatrix} & \quad\!\!\!\pi\mbox{-phase}
\end{cases}~~.
\label{SigBZero}
\end{equation}
The situation at zero magnetic fields is thus more complex for perturbative methods.

\begin{figure*}[ht]
\begin{center}
\includegraphics[width=0.8\textwidth]{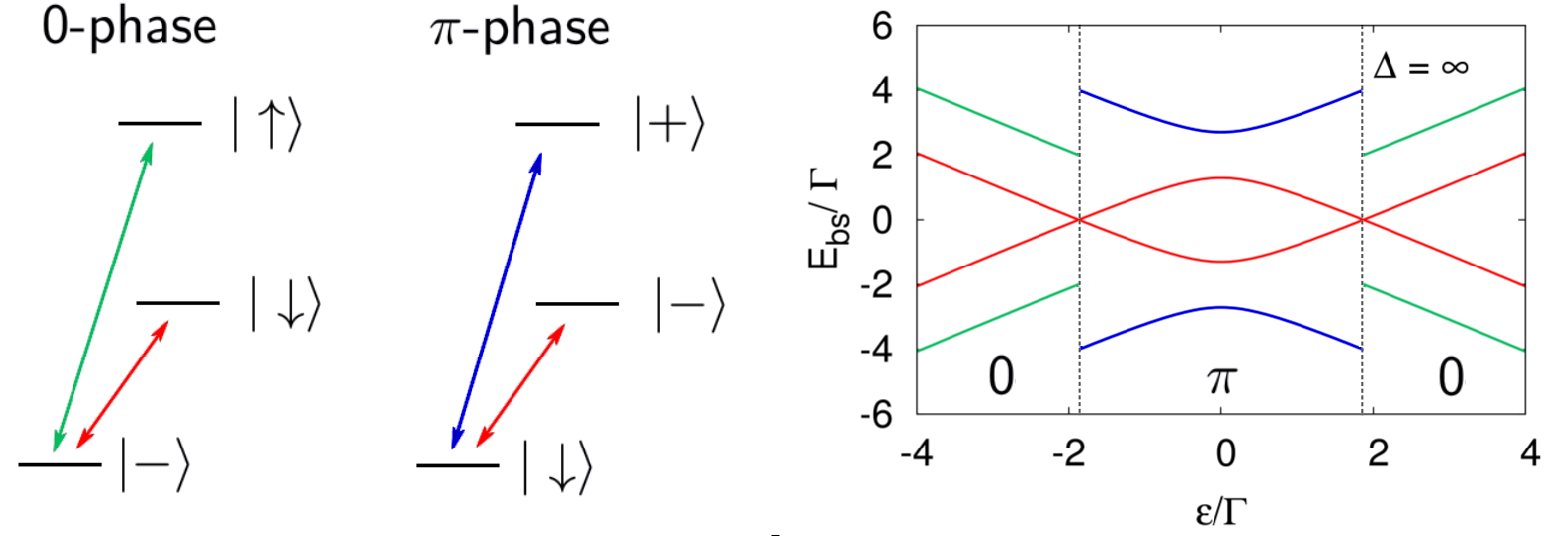}
\caption{ The visible Andreev bound states and the corresponding transitions in
and out of the ground states in the $0$- and $\pi$-phase for $U=2\Gamma$,
$B=0.7\Gamma$, $\phi=\pi/2$ and $\Gamma_L=\Gamma_R=\Gamma/2$.}
\label{ABS_ATLI}
\end{center}
\end{figure*}

\begin{figure}[b]
\begin{center}
\includegraphics[width=0.85\columnwidth]{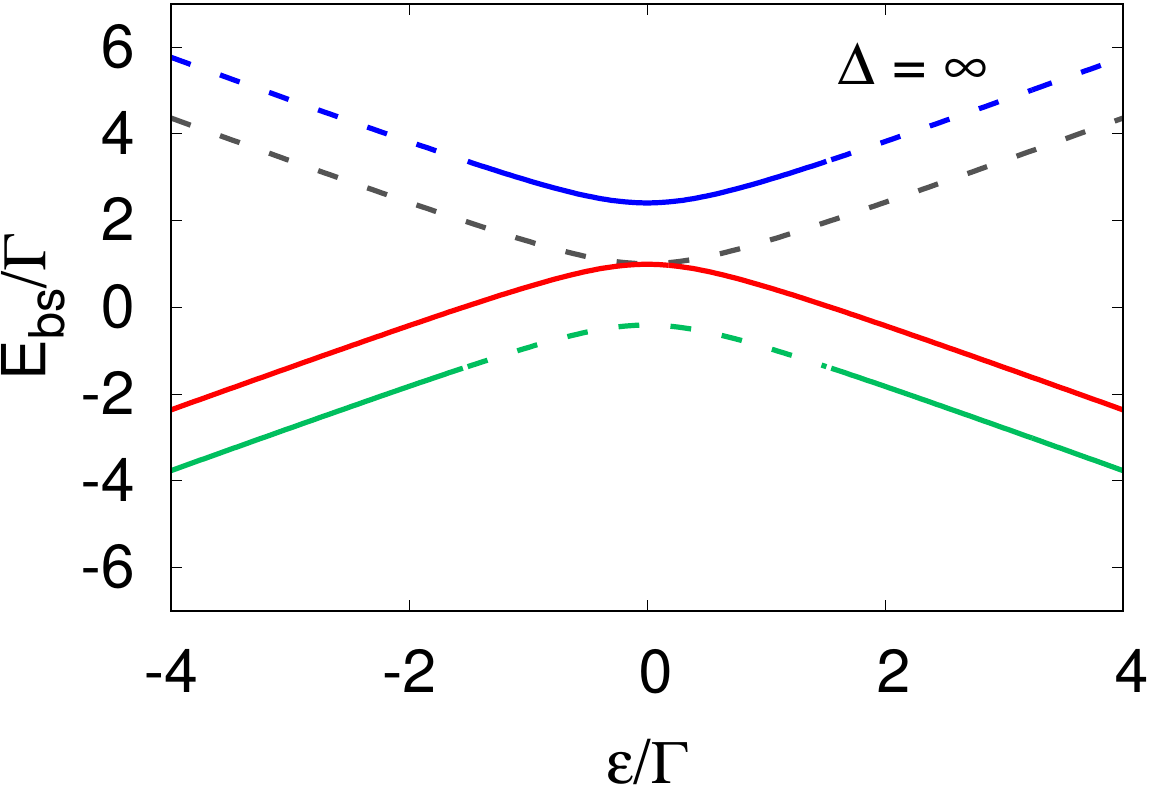}
\caption{The Andreev transition energies $a_\uparrow$, $a_\downarrow$,
$b_\uparrow$ and $b_\downarrow$ (bottom to top) for the large-gap limit as a function of the
on-site energy $\epsilon$ and $U=2\Gamma$, $B=0.7\Gamma$, $\phi=\pi/2$ and
$\Gamma_L=\Gamma_R=\Gamma/2$. Solid lines correspond to regions of non-vanishing
weight, while dotted lines denote a vanishing weight. Note that the
contributions $-a_\uparrow$, $-a_\downarrow$, 
$-b_\uparrow$ and $-b_\downarrow$ from the symmetric ABS have not been drawn 
here for clarity.}
\label{AndreevEnergies}
\end{center}
\end{figure}

To get a more physical understanding of the Andreev bound state energies, we
refer again to the Lehmann representation of the Green's function in the atomic
limit. Here, the poles can be identified as one-electron transitions between the
eigenstates
$\{|-\rangle,|+\rangle\} \leftrightarrow \{|\uparrow\rangle, |\downarrow\rangle\}$.
The possible transition energies are thus
\begin{subequations}
\begin{align}
a_\sigma = E_- - \sigma B \\
b_\sigma = E_+ - \sigma B
\end{align} 
\end{subequations}
and their negative values respectively. The corresponding weights of the Andreev
bound states are summarized in Table \ref{TabWeight} (see
App.~\ref{app:LargeGap} for details) for both phases in the case of finite
magnetic field $B>0$. The expressions $a_\sigma$ and $b_\sigma$ are plotted in
\cfg{AndreevEnergies} as a function of the on-site energy $\epsilon$ and for
$U=2\Gamma$, $B=0.7\Gamma$ and $\phi=\pi/2$. Here, solid lines were chosen
whenever the corresponding weight is non-vanishing, and dashed line
are associated to zero weight, thus to a non-visible transition.
\begin{table}
\begin{minipage}{0.9\columnwidth}
\begin{center}
\begin{tabular}{c|c|c|c}
\multicolumn{4}{c}{$0$-phase} \\
$E_{\rm bs}$ & Transition & $w$ & $w_\Delta$ \\
\hline
$\pm a_\uparrow$ & $|\uparrow\rangle \leftrightarrow |-\rangle$ & $|v|^2,|u|^2$ & $0,-u^*v$ \\
\hline
$\pm a_\downarrow$ & $|\downarrow\rangle \leftrightarrow |-\rangle$ & $|v|^2,|u|^2$ & $u^*v,0$ \\
\hline
$\pm b_\uparrow$ & $|\uparrow\rangle \leftrightarrow |+\rangle$ & 0 & 0\\
\hline
$\pm b_\downarrow$ & $|\downarrow\rangle \leftrightarrow |+\rangle$ & 0  & 0
\end{tabular}
\end{center}
\end{minipage}

\vspace{.25cm}

\begin{minipage}{0.9\columnwidth}
\begin{center}
\begin{tabular}{c|c|c|c}
\multicolumn{4}{c}{$\pi$-phase} \\
$E_{\rm bs}$ & Transition & $w$ & $w_\Delta$ \\
\hline
$\pm a_\uparrow$ & $|\uparrow\rangle \leftrightarrow |-\rangle$ & 0 & 0\\
\hline
$\pm a_\downarrow$ & $|\downarrow\rangle \leftrightarrow |-\rangle$ & $|v|^2,|u|^2$ & $u^*v,0$\\
\hline
$\pm b_\uparrow$ & $|\uparrow\rangle \leftrightarrow |+\rangle$ & 0 & 0\\
\hline
$\pm b_\downarrow$ & $|\downarrow\rangle \leftrightarrow |+\rangle$ & $|u|^2,|v|^2$ & $-u^*v,0$\\
\end{tabular}
\end{center}
\end{minipage}
\caption{Spectral weights and anomalous weights of the Andreev bound states evaluated %(a) 
for the $0$-phase and %(b) 
for the $\pi$-phase, with the associated transitions.} 
\label{TabWeight}
\end{table}

Let us now clarify a few important points that will allow for a deeper
understanding of the ABS even for the case of finite gap.  First we want to
point out that at finite magnetic field exactly two bound states (four,
including their symmetric partners) have a non-vanishing weight, independently
whether the ground state is magnetic or not.
The energies of the inner bound state pair are given $\pm a_\downarrow$ in both
phases, and can thus be tracked continuously across the phase transition.
Further, as the requirement for the level crossing phase-transition is given by
$E_\downarrow=E_-$ and thus $a_\downarrow=0$, the inner bound state will always
cross the chemical potential at the point of the phase transition, while the
outer bound state pair experiences a jump in energy.  While in the $0$-phase the
outer bound-state pair has energies $\pm a_\uparrow$, their energies change to
$\pm b_\downarrow$ in the $\pi$-phase.  This behavior is depicted in
\cfg{ABS_ATLI} for the case of a varying level position $\epsilon$. Here and
in the following we show the inner bound states $\pm a_\downarrow$ in red, while
$\pm a_\uparrow$ is shown in green and $\pm b_\downarrow$ in blue.

We finally consider the Josephson current in the large gap limit for %focusing on the case of 
a non-vanishing magnetic field. 
The total current is most straightforwardly calculated by the derivative of the
ground state energy $E_{\rm GS}(\phi)$
\begin{equation}
J = 2 \partial_{\phi} E_{\rm GS}(\phi).
\end{equation}
In the $\pi$-phase, the ground state energy does not exhibit any
$\phi$-dependence, leading to a vanishing Josephson current.
In the $0$-phase, the current is given by
\begin{equation}
\label{jcurr}
J = -2\partial_\phi E_{\phi} = 2\Gamma_L\Gamma_R\frac{\sin\phi}{E_{\phi}}.
\end{equation}
It is instructive to determine the %individual 
contribution of each bound state 
to the total Josephson current. 
In the limit $\Delta\rightarrow\infty$ Eq.~(\ref{JCBS}) yields
\begin{equation}
\langle J_{E_{\rm bs}} \rangle = - 2 \Gamma_L \Im\mathrm{m} \left [e^{i\phi/2} w_\Delta(E_{\rm bs})^*
\right ] \mbox{sgn}(E_{\rm bs}),
\end{equation}
Since the spectrum on the dot consists only of the bound states, we get
no continuum contribution to the total Josephson current. 
Recalling that $u^*v = \Gamma_{\phi}/(2E_{\phi})$, the result for the $0$-phase is
\begin{equation}
\langle J_{-a_\uparrow} \rangle = \langle J_{a_\downarrow} \rangle = \Gamma_L \Gamma_R \frac{\sin\phi}{E_{\phi}},
\end{equation}
adding up to the %expected 
total Josephson current~\eqref{jcurr}.
In the $\pi$-phase the contributions are
\begin{equation}
\langle J_{b_\downarrow} \rangle = -\langle J_{a_\downarrow} \rangle = \Gamma_L \Gamma_R \frac{\sin\phi}{E_{\phi}},
\end{equation}
leading to a vanishing Josephson current, as expected. 
Having identified the transitions associated to the different bound state energies %with specific transitions 
(see Table \ref{TabWeight}), we can interpret the 
corresponding Josephson current contribution as a measure for the relevance of the virtual intermediate state in the Cooper pair transport process. 
It is also interesting to note that the magnitude 
of the current in the $0$-phase does not depend on the magnetic field at large gap, an artifact of this limit.

\section{Methods}
\label{sec:methods}

We here briefly review two complementary approaches that are able to tackle the problem
of superconducting quantum dots in presence of both a finite Coulomb interaction and a finite gap:
the static fRG and the SCABS approximation.
In the description of their implementation, we focus on the aspects specific to the extension to finite magnetic fields.

\subsection{Static functional renormalization group}

The fRG~\cite{Metzner2012,Salmhofer1999} is based on Wilson's general RG idea
for interacting many-body systems. By introducing a scale-dependence into the
non-interacting Green's function one can derive an exact functional flow
equation, that describes the gradual evolution of the effective action, that is,
the generating functional of the one-particle irreducible vertex functions, 
as the scale is changed.
While the action at the final scale is the one of the system in question, 
we only require the initial action to be exactly solvable, giving rise to a 
large freedom in the choice of the initial conditions\cite{Wentzell2015}.
Expanding this functional flow equation
in powers of the external sources yields an exact but infinite hierarchy of flow equations for
the $n$-particle vertex functions.  In practical implementations, however, this
hierarchy has to be truncated at a given order. This truncation is commonly
performed at the two-particle level, and yields a set of flow equations for the
self-energy and the two-particle vertex functions.

We here use the fRG implementation for superconducting quantum dots formulated
on the Matsubara axis~\cite{Karrasch2006,Metzner2012} (see
Ref.~\onlinecite{Rentrop2014} for the extension to real-time Keldysh space)
assuming that the self-energy and the two-particle vertex are both static. The underlying
approximations are devised for weak to intermediate Coulomb interaction
strengths and arbitrary gap, and have been checked by comparing with NRG data. 

At zero temperature, we use a frequency cutoff of the form 
\begin{align}
G_0^\Lambda = \Theta(|\omega| - \Lambda) G_0.
\end{align}
while the Green function at a given scale is determined by means of the 
Dyson equation $G^\Lambda = \left[(G_0^{\Lambda})^{-1} - \Sigma^\Lambda \right]^{-1}$.
In the static approximation, the self-energy contains %only 
three frequency-independent elements %components?
\begin{align}
\Sigma^\Lambda(i\omega) = \begin{pmatrix} \Sigma^\Lambda_1 &
\Sigma^\Lambda_\Delta \\ {\Sigma^{\Lambda}_{\Delta}}^* & \Sigma^\Lambda_2
\end{pmatrix},
\end{align}
while the static two-particle vertex is determined by a single renormalized
Coulomb interaction $U^\Lambda$.
Note that the static terms $\Sigma^\Lambda_{1}$ and $\Sigma^\Lambda_{2}$ effectively renormalize 
the on-site energy and magnetic field.
Introducing flowing effective physical parameters
\begin{equation}
\epsilon^\Lambda = \epsilon + \frac{1}{2} \left(\Sigma^\Lambda_1 -
\Sigma^\Lambda_2 \right), \qquad B^\Lambda = B + \frac{1}{2}
\left(\Sigma^\Lambda_1 + \Sigma^\Lambda_2 \right),
\end{equation}
the Green's function reads
\begin{equation}
\begin{split}
G^\Lambda(i\omega) = \frac{1}{D(i\omega)} \begin{pmatrix} i\tilde{\omega} +
\epsilon^\Lambda - B^\Lambda  & -\tilde{\Delta} +\Sigma^\Lambda_\Delta \\
-\tilde{\Delta}^* + {\Sigma^\Lambda_\Delta}^* & i\tilde{\omega}   -
\epsilon^\Lambda - B^\Lambda \end{pmatrix},
\end{split}
\end{equation}
with the determinant
\begin{align}
\label{det}
  {D(i\omega)} = \left( i\tilde{\omega} - \epsilon^\Lambda - B^\Lambda \right)
\left(i\tilde{\omega} 
  + \epsilon^\Lambda - B^\Lambda \right)
  - |\tilde{\Delta} - \Sigma^\Lambda_\Delta|^2.
\end{align}
The explicit flow equations for the effective %renormalized 
parameters read
\begin{widetext}
\begin{subequations}
\begin{align}
\partial_\Lambda \epsilon^\Lambda &= \frac{U^\Lambda \epsilon^\Lambda}{\pi
|D(i\Lambda)|^2}  \left[\tilde{\omega}^2 + \left(\epsilon^\Lambda \right)^2 -
\left(B^\Lambda \right)^2 + |\tilde{\Delta}-\Sigma^\Lambda_\Delta|^2\right]_{\omega=\Lambda},\\
\partial_\Lambda B^\Lambda &= \frac{U^\Lambda B^\Lambda}{\pi |D(i\Lambda)|^2}
\left[\tilde{\omega}^2 - \left(\epsilon^\Lambda \right)^2 + \left(B^\Lambda
\right)^2 + |\tilde{\Delta}-\Sigma^\Lambda_\Delta|^2\right]_{\omega=\Lambda},\\
\partial_\Lambda \Sigma^\Lambda_\Delta &= \frac{U^\Lambda (\Sigma^\Lambda_\Delta
- \tilde{\Delta})}{\pi |D(i\Lambda)|^2}  \left[\tilde{\omega}^2 +
  \left(\epsilon^\Lambda \right)^2 - \left(B^\Lambda \right)^2 +
|\tilde{\Delta}-\Sigma^\Lambda_\Delta|^2\right]_{\omega=\Lambda},
\end{align}
\end{subequations}
\end{widetext}
and
\begin{align}
\quad \partial_\Lambda U^\Lambda = 2\pi \left[ \left(\partial_\Lambda
B^\Lambda\right)^2 - \left(\partial_\Lambda \epsilon^\Lambda\right)^2 +
|\partial_\Lambda \Sigma^\Lambda_\Delta|^2\right]_{\omega=\Lambda}
\end{align}
for the two-particle vertex, with the initial conditions
\begin{subequations}
\begin{align}
\epsilon^{\Lambda=\infty} &= \epsilon, \qquad B^{\Lambda=\infty} = B,\\
\Sigma_\Delta^{\Lambda=\infty} &= 0, \qquad U^{\Lambda=\infty} =U.
\end{align}
\end{subequations}
This set of ordinary differential equations is then integrated numerically from 
$\Lambda/\Gamma=10^6$ to $\Lambda/\Gamma=10^{-6}$ using a Runge-Kutta solver.
An example for the evolution of the renormalized parameters during the flow is shown in \cfg{FLOW_EPS}.
\begin{figure}[t]
\begin{center}
\includegraphics[width=0.85\columnwidth]{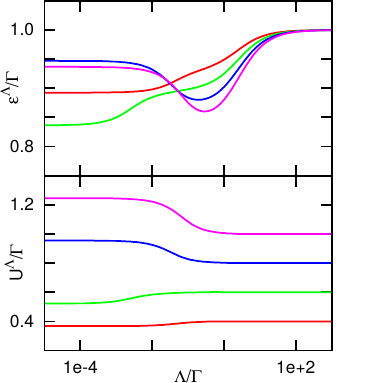}
\caption{Flow of the renormalized on-site energy $\epsilon^\Lambda$ and the 
effective interaction $U^\Lambda$ for $\Delta=\Gamma$, $\epsilon=\Gamma$, 
$B=\Gamma$, $\phi=\pi/2$, $\Gamma_L=\Gamma_R=\Gamma/2$ and different values 
of $U$. $U=0.6\Gamma$ is close to the phase transition and the flow converges 
at a lower energy scale. Note that the interaction is effectively reduced in 
the $0$-phase, while an enhancement is observed in the $\pi$-phase.}
\label{FLOW_EPS}
\end{center}
\end{figure}

Introducing the notation
\begin{subequations}
\begin{align}
\epsilon^{\Lambda=0} &= \epsilon_r, \qquad \hspace{0.17cm}B^{\Lambda=0} = B_r,\\
\Sigma_\Delta^{\Lambda=0} &= \Sigma_\Delta, \qquad U^{\Lambda=0} =U_r
\end{align}
\end{subequations}
for the renormalized values at the end of the flow,  
the poles of the Green's function are determined by finding the roots of its determinant \eqref{det}, e.g. by solving
\begin{align}
\left(\tilde{\omega} - \epsilon_r - B_r \right)\left(\tilde{\omega}
+ \epsilon_r - B_r  \right)
- |\tilde{\Delta} - \Sigma_{\Delta}|^2=0.
\label{Determ}
\end{align}
The spectral weights of the associated ABS are then calculated according to Eqs.~(\ref{NormWeight}) and (\ref{AnWeight}).

\subsection{Self-consistent Andreev bound state theory}
\label{subsec:SCABS}

This alternative approach focuses again on effective Andreev levels,
but, instead of a scheme based on a renormalized perturbative expansion in
the Coulomb interaction, rather considers the infinite gap limit as a 
starting point for a perturbative treatment. 
The clear advantage here is that the $0$ to $\pi$
transition is already captured at $\Delta=\infty$, and thus the method
should be able to describe both phases on an equal footing.
For $\Delta=\infty$, we have previously calculated the one-particle 
energy levels, $E_{\sigma}^0=\sigma B$, and the BCS-like levels, 
$E_{\pm}^0=U/2 \pm \sqrt{\epsilon^2  + |\Gamma_{\phi}|^2}$. Note that we have added
an additional superscript 0, to denote that these are the uncorrected energies at 
infinite gap. Further, all following derivations will be considering the general case of a finite
bandwidth $2D$, which requires the introduction of the generalized hybridization function 
$\Gamma_{\phi}(i\omega) = \frac{2}{\pi} \arctan\left(\frac{D}{\sqrt{\Delta^2-(i\omega)^2}}\right) \sum_\alpha \Gamma_\alpha e^{i\phi_\alpha}$.
In the following, $\Gamma_{\phi}=\Gamma_{\phi}(0)$. 

Straightforward calculations detailed in App.~\ref{app:SCABS} give
the perturbative correction at lowest order
\begin{widetext}
\begin{subequations}
\begin{eqnarray}
\nonumber
\delta E_\sigma \!&=&\! -t^2\sum_{\vec{k}}{\left[\frac{1}{E_{\vec{k}} 
+ (E_+^0-E_{\sigma}^0)} + \frac{1}{E_{\vec{k}} + (E_-^0-E_{\sigma}^0)} 
+ \frac{2\Delta}{E_{\vec{k}}}uv\left|\cos\frac{\phi}{2}\right|\left(\frac{1}{E_{\vec{k}} 
+ (E_+^0-E_{\sigma}^0)} - \frac{1}{E_{\vec{k}} + (E_-^0-E_{\sigma}^0)}\right)\right]}\\
\label{eq:Es}\\
\label{eq:E+}
\delta E_+ \!&=&\! -t^2\sum_{\vec{k},\sigma}{\left(\frac{1}{E_{\vec{k}}
-(E_+^0-E_{\sigma}^0)}-\frac{2\Delta}{E_{\vec{k}}}uv\left|\cos\frac{\phi}{2}\right|\frac{1}{E_{\vec{k}}
-(E_+^0-E_{\sigma}^0)}\right)}-2|\Gamma_{\phi}|uv\\
\label{eq:E-}
\delta E_- \!&=&\! -t^2\sum_{\vec{k},\sigma}{\left(\frac{1}{E_{\vec{k}}
-(E_-^0-E_{\sigma}^0)}+\frac{2\Delta}{E_{\vec{k}}}uv\left|\cos\frac{\phi}{2}\right|\frac{1}{E_{\vec{k}}
-(E_-^0-E_{\sigma}^0)}\right)}+2|\Gamma_{\phi}|uv .
\end{eqnarray}
\end{subequations}
\end{widetext}
with the quasiparticle energy $E_{\vec{k}} = \sqrt{{\epsilon_{\vec{k}}}^2+{\Delta}^2}$.
These expressions generalize the results of Ref.~\onlinecite{Meng2009a} to the
case of finite magnetic field.

At finite $\Delta$, the self-consistent perturbative approach considered in
Ref.~\onlinecite{Meng2009a} can be generalized to the spinful case. 
In order to write self-consistent equations for the corrections to the Andreev
transitions, $\delta a_\sigma = \delta E_- - \delta E_{\sigma} = a_\sigma - a_\sigma^0$ and
$\delta b_\sigma = \delta E_+ - \delta E_{\sigma} = b_\sigma - b_\sigma^0$, 
one must analyze carefully the singularities appearing in their respective
expressions:
\begin{widetext}
\begin{subequations}
\begin{align}
\label{eq:deltaa}
\delta a_\sigma & = -\frac{\Gamma}{\pi}\int_{0}^{D}d\epsilon \, 
\Bigg[\sum_{\sigma'}\frac{1}{E-a_{\sigma'}^0}-\frac{1}{E+b_\sigma^0}
-\frac{1}{E+a_\sigma^0}
+\frac{2\Delta}{E}uv\left|\cos\frac{\phi}{2}\right|
\Bigg(\sum_{\sigma'}\frac{1}{E-a_{\sigma'}^0}
-\frac{1}{E+b_\sigma^0}
+\frac{1}{E+a_\sigma^0}\Bigg)\Bigg] +2|\Gamma_{\phi}|uv,\\
\label{eq:deltab}
\delta b_\sigma & = -\frac{\Gamma}{\pi}\int_{0}^{D}d\epsilon \, 
\Bigg[\sum_{\sigma'}\frac{1}{E-b_{\sigma'}^0}
-\frac{1}{E+b_\sigma^0}-\frac{1}{E+a_\sigma^0}
+\frac{2\Delta}{E}uv\left|\cos\frac{\phi}{2}\right|
\Bigg(\sum_{\sigma'}\frac{-1}{E-b_{\sigma'}^0}
-\frac{1}{E+b_\sigma^0} +\frac{1}{E+a_\sigma^0}\Bigg)\Bigg] -2|\Gamma_{\phi}|uv \text{.}
\end{align}
\end{subequations}
\end{widetext}
Recall that $E = \sqrt{\epsilon^2 + \Delta^2}$, such that singularities appear indeed 
whenever a one-particle transition on the dot becomes
comparable to the minimum quasiparticle energy given by the gap $\Delta$. A 
first important observation is that the singularities tend to cancel out together
for the outer bound state correction $\delta b_\sigma$, which implies that these 
states become part of the continuum for small enough $\Delta$. One can thus focus on
analyzing the singularities related to the inner bound states $a_\sigma$, which
originate from the denominators in $1/(E\pm a_\sigma^0)$. The physics here is
simply an effect of level repulsion from the continuum whenever the bound state
approach the gap edges. In the case $a_\sigma^0>0$, which occurs typically in
the regime of strong correlations $U\gg \Gamma$, only the denominators in 
$1/(E - a_\sigma^0)$ are singular. This leads to a downward renormalization
of the bound state energy $a_\sigma$ compared to the bare value $a_\sigma^0$.
Conversely, an upward renormalization of the bound state occurs when 
$a_\sigma^0<0$, since the denominators $1/(E + a_\sigma^0)$ provide then the main 
contribution. We can thus renormalize in a self-consistent way the inner Andreev
bound states according to
\begin{widetext}
\begin{align}
\label{eq:a}
\delta a_\sigma  = &-\frac{\Gamma}{\pi}\int_{0}^{D}d\epsilon \, 
\Bigg[\sum_{\sigma'}\frac{1}{E-a_{\sigma'}^0-\Theta[-\delta a_{\sigma'}]
\delta a_{\sigma'}} -\frac{1}{E+b_\sigma^0}
-\frac{1}{E+a_\sigma^0}\nonumber\\
&+\frac{2\Delta}{E}uv\left|\cos\frac{\phi}{2}\right|
\Bigg(\sum_{\sigma'}\frac{1}{E-a_{\sigma'}^0-\Theta[-\delta a_{\sigma'}]\delta a_{\sigma'}}
-\frac{1}{E+b_\sigma^0}
 +\frac{1}{E+a_\sigma^0+\Theta[\delta a_{\sigma'}]\delta a_{\sigma'}}\Bigg)\Bigg] +2|\Gamma_{\phi}|uv, %\\
 %\label{eq:b}
 %+\frac{1}{E+a_\sigma^0}\Bigg)\Bigg] -2|\Gamma_{\phi}|uv.
\end{align}
\end{widetext}
and correspondingly for b.
\begin{figure*}[ht]
\begin{center}
\includegraphics[width=0.8\textwidth]{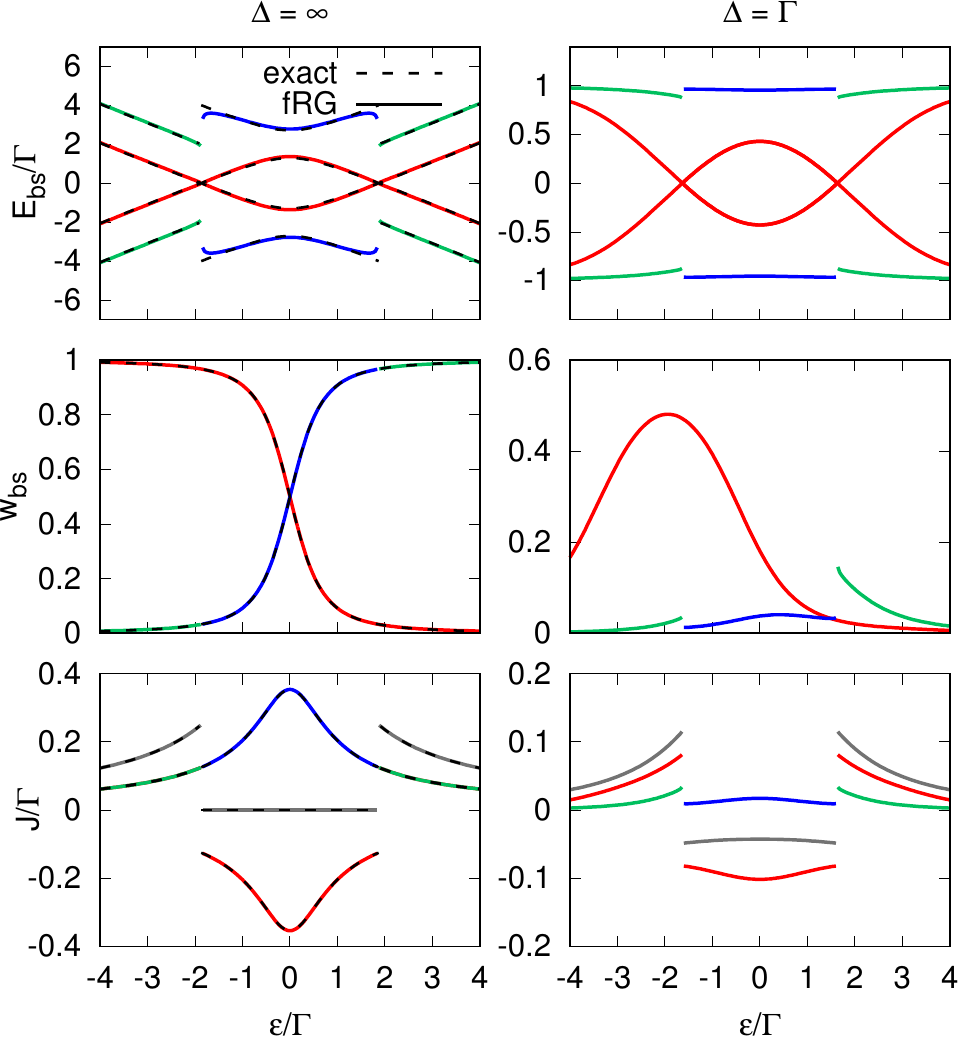}
\caption{Bound state energies (upper panels), the corresponding weights (middle
panels) as well as the bound-state resolved Josephson currents (lower panels) 
defined in \ceq{JCBS} as a function of
the on-site energy $\epsilon$. The calculation is shown for the large-gap
situation (left column) and a finite gap $\Delta=\Gamma$ (right column), with
$U=2\Gamma$, $\phi=\pi/2$ and $B=0.7\Gamma$ in all cases. Solid lines show fRG
results, while dotted lines denote the exact expressions for $\Delta=\infty$. 
The weights shown here correspond to the bound state energies $a_\uparrow$, 
$a_\downarrow$ and $-b_\downarrow$ associated with their respective colors 
(compare \cfg{AndreevEnergies}). The gray lines denote the total Josephson current. }
\label{fRGATLI}
\end{center}
\end{figure*}
Note the presence here of $\Theta$-functions that account for respective
downward and upward renormalization, as discussed above.
We thus find that $\delta a_{\sigma}$ depends on both $\delta
a_{\uparrow}$ and $\delta a_{\downarrow}$, such that one has to solve a coupled
set of self-consistent equations for $\delta a_{\sigma}$ (and similarly for
$\delta b_{\sigma}$). These equations can, however, be decoupled, since $\delta
a_{\uparrow}-\delta a_{\downarrow}$ is a constant that does not depend on either
$\delta a_{\sigma}$ (and again similarly for $\delta b_{\sigma}$, which is not
written here). This simple procedure does not provide any information on the weights
of the ABS, in contrast to the fRG approach of the previous section. The
understanding of the allowed transitions can nevertheless be gathered from the
atomic limit.

\section{Results}
\label{sec:results}

For the results in the following we will focus on the case of symmetric coupling
$\Gamma_L = \Gamma_R$ as the physics of the system does not differ from the
general case.  We will first describe how the case of finite gap is linked to
the solution in the large-gap limit in order to understand in more detail the
effect of a local magnetic field on the spectrum.  This will be followed by a
detailed comparison between the fRG and the SCABS approximation, and further by a brief
benchmark against available NRG results~\cite{Bauer2007}. To conclude our
study, we will give a small outlook towards transport calculations that are closer 
to actual spectroscopic experimental setups.

\subsection{From large to small gaps using fRG}

While the previously introduced SCABS approximation includes the exact large-gap limit
solution by construction, this does not hold for the fRG. This allows us to
benchmark fRG calculations of the Andreev bound states performed for a large gap
value (e.g. $10^6\Gamma$) against the exact expressions presented previously.
This comparison is shown in the left panel of \cfg{fRGATLI}, which shows the
Andreev bound state energies (upper panels), the corresponding spectral weights 
(middle panels) as well as the bound-state resolved Josephson current (lower
panels) as a function of the level-position $\epsilon$ for $U=2\Gamma$, $\phi=\pi/2$. The
dashed line indicates the exact solution in the large-gap limit, while solid
lines denote the corresponding fRG data. Bound state colors are chosen as
previously introduced.  We find an excellent agreement of the fRG data with the
exact solution, not only for the ABS, but also for their weights as well as the
Josephson currents. Small deviations can be found in the $\pi$-phase for the outer
bound states $\pm b_\downarrow$ (blue), specifically close to the phase
transition.

The corresponding fRG data for the same set of parameters but now a finite gap
$\Delta=\Gamma$ is shown in the right panels of \cfg{fRGATLI}. While the
qualitative behavior of the ABS is similar, we find that, due to the repulsion
from the gap-edge, the overall structure is squeezed in the process of closing
the gap from large to small values. In particular the outer bound states are
strongly deformed due to this process.  This is also mirrored in the change of
the spectral weight, as the ABS tend to loose more weight the closer they are to
the gap edge. In fact, for sufficiently small gap, the outer bound state pair
can be absorbed completely into the continuum part of the spectrum.
As the gap is lowered, we also find a non-vanishing Josephson
current (gray) in $\pi$-phase. Further it is interesting to note, that the
bound-state contributions no longer add up to the total Josephson current
(which we will study in what follows), as the continuous part of the
DOS has now a non-vanishing contribution to the Josephson current.

\subsection{Magnetic field effects}
	
\begin{figure}[t]
\begin{center}
\includegraphics[width=0.95\columnwidth]{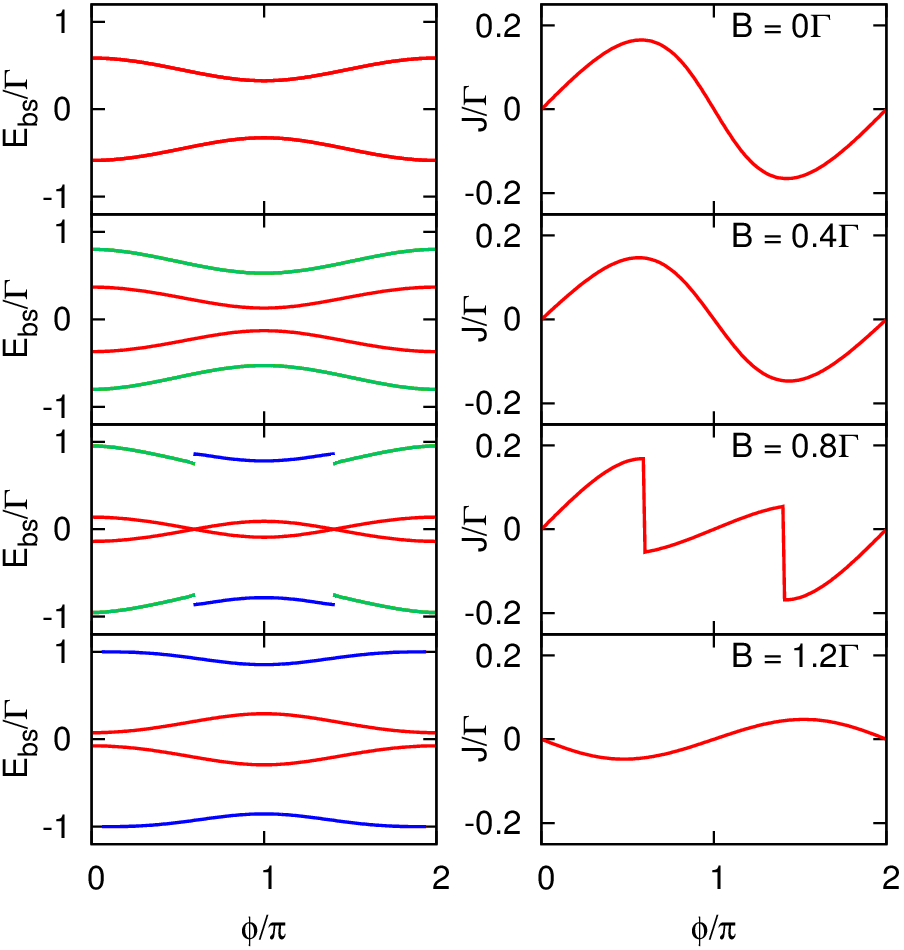}
\caption{Evolution of the Andreev bound states and Josephson current with $\phi$ as obtained from fRG for $\epsilon=\Gamma$, $\Delta=\Gamma$, $U=\Gamma$ and different values of $B$.}
\label{ABS_B_dependence}
\end{center}
\end{figure}

We here discuss how the magnetic field alters the ABS in the $0$-phase and by
this drives the phase-transition. Figure \ref{ABS_B_dependence} shows the ABS (left)
and Josephson current (right) obtained from fRG, as a function of the 
phase-difference $\phi$ for
$\epsilon=\Gamma$, $\Delta=\Gamma$, $U=\Gamma$ and different values of $B$. In
the absence of a magnetic field (upper panel), the system is in the $0$-phase for
the whole $\phi$-range, and the visible ABS $a_\uparrow$ and $a_\downarrow$ are
equal. Accordingly, the Josephson current shows the typical sinusoidal behavior
without a jump. 

For a small magnetic field ($B=0.4\Gamma$) the $0$-phase is still the most
stable, but the bound states $a_\uparrow$ and $a_\downarrow$ can now be clearly
distinguished due to the Zeeman splitting. The corresponding Josephson is
just mildly reduced as a consequence.

When increasing the magnetic field further ($B=0.8\Gamma$), the inner bound
states $\pm a_\downarrow$ will cross the chemical potential for $\phi$ close to
$\pi$, thus inducing the phase transition for a finite $\phi$-range.  In this
window, the visible outer bound-state changes to $\pm b_\downarrow$. A Zeeman splitting
is thus no longer directly visible in this part of the spectrum. As expected,
the change of the ground state is accompanied by a sign-reversal in the
Josephson current.

For even larger values of the magnetic field ($B=1.2\Gamma$), the inner bound
states will completely cross the chemical potential, inducing the $\pi$-phase
for the whole $\phi$-range. Accordingly, the Josephson current completely
inverts its sign.

\subsection{Comparison between fRG and the SCABS approximation}

\begin{figure*}[t!]
\begin{center}
\includegraphics[width=0.8\textwidth]{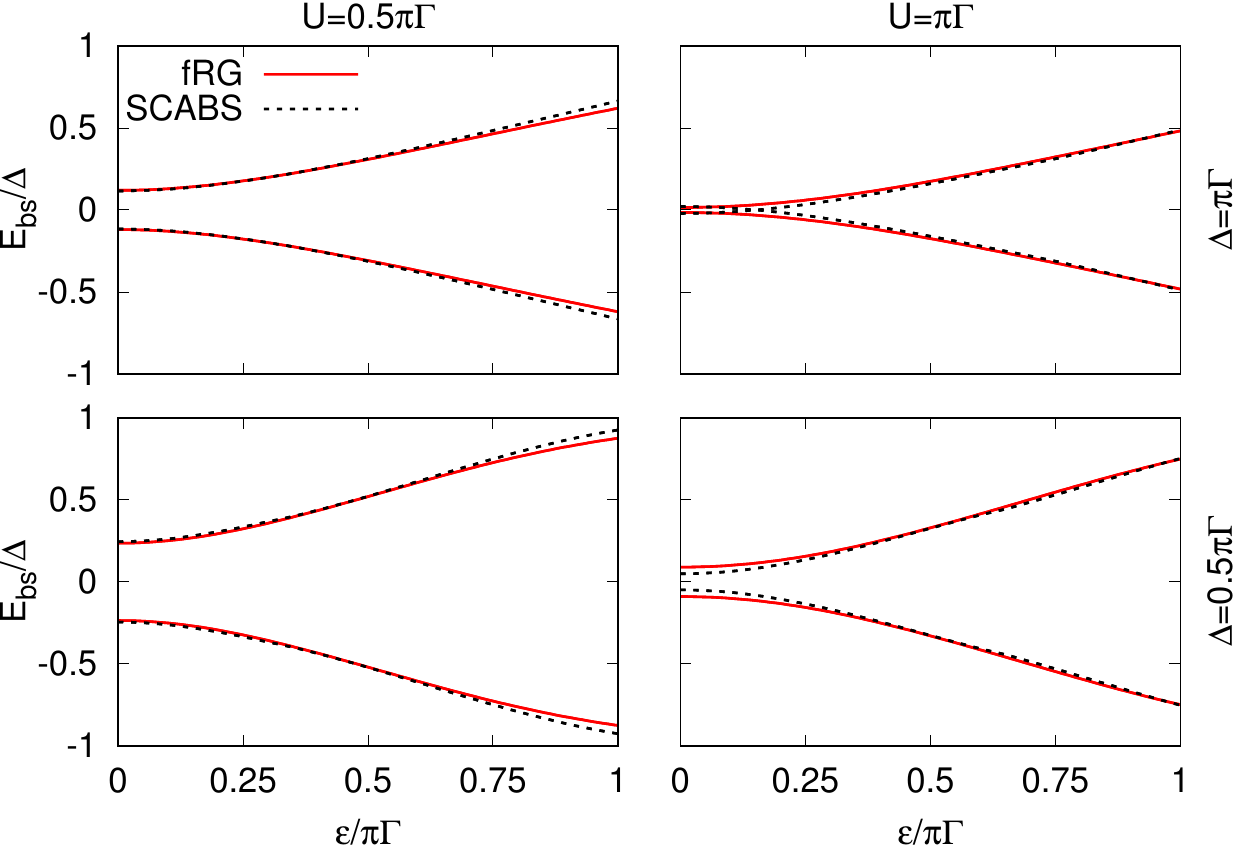}
\caption{Bound state energies calculated with fRG (full lines) and SCABS approximation
(dashed lines) as a function of $\epsilon$ for $B=0$, $\phi=0$ and different 
values of $U$ and $\Delta$.}
\label{fig_varEps}
\end{center}
\vspace{.3cm}
\begin{center}
\includegraphics[width=0.8\textwidth]{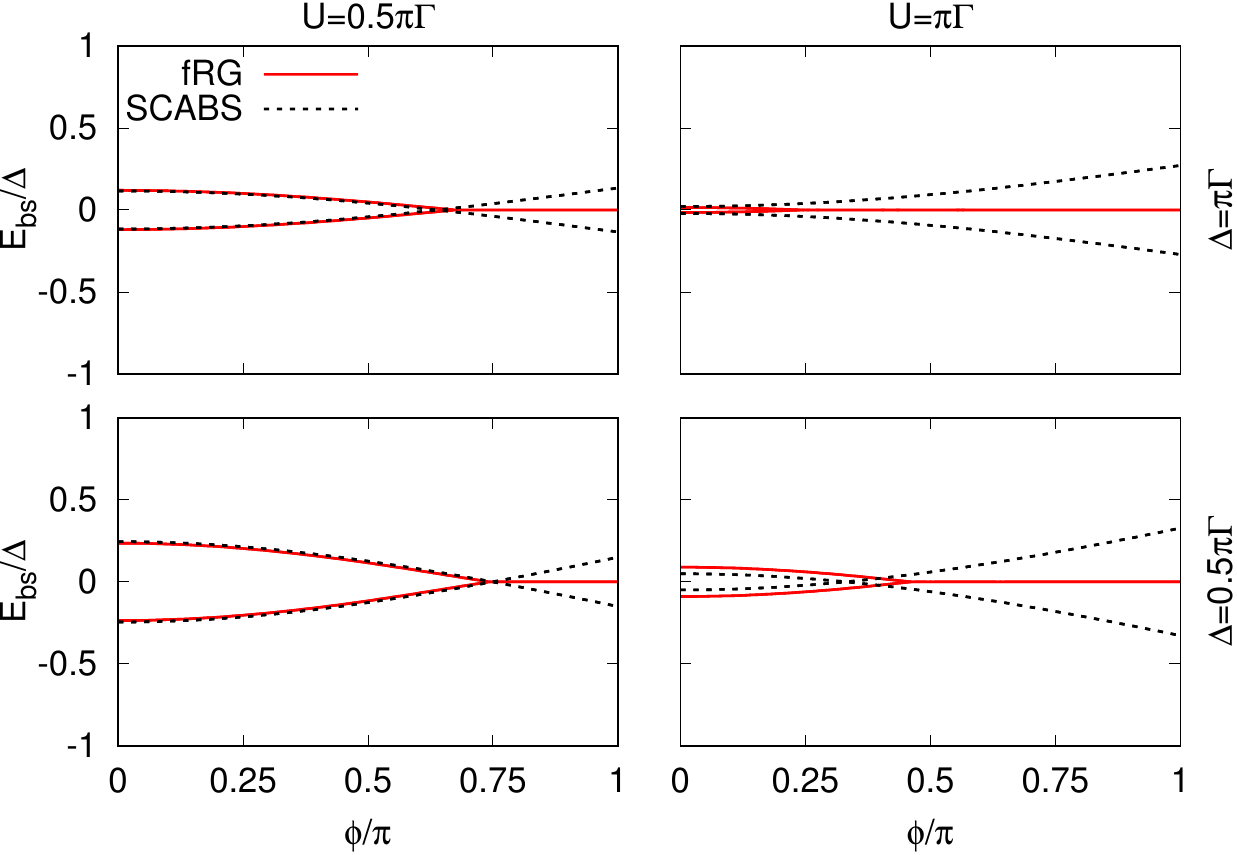}
\caption{Bound state energies calculated with fRG (full lines) and SCABS approximation 
(dashed lines) as a function of $\phi$ for $B=0$, $\epsilon=0$ and different values of $U$ and $\Delta$. }
\label{fig_varPhi}
\end{center}
\end{figure*}

\begin{figure*}[t]
\begin{center}
\includegraphics[width=0.8\textwidth]{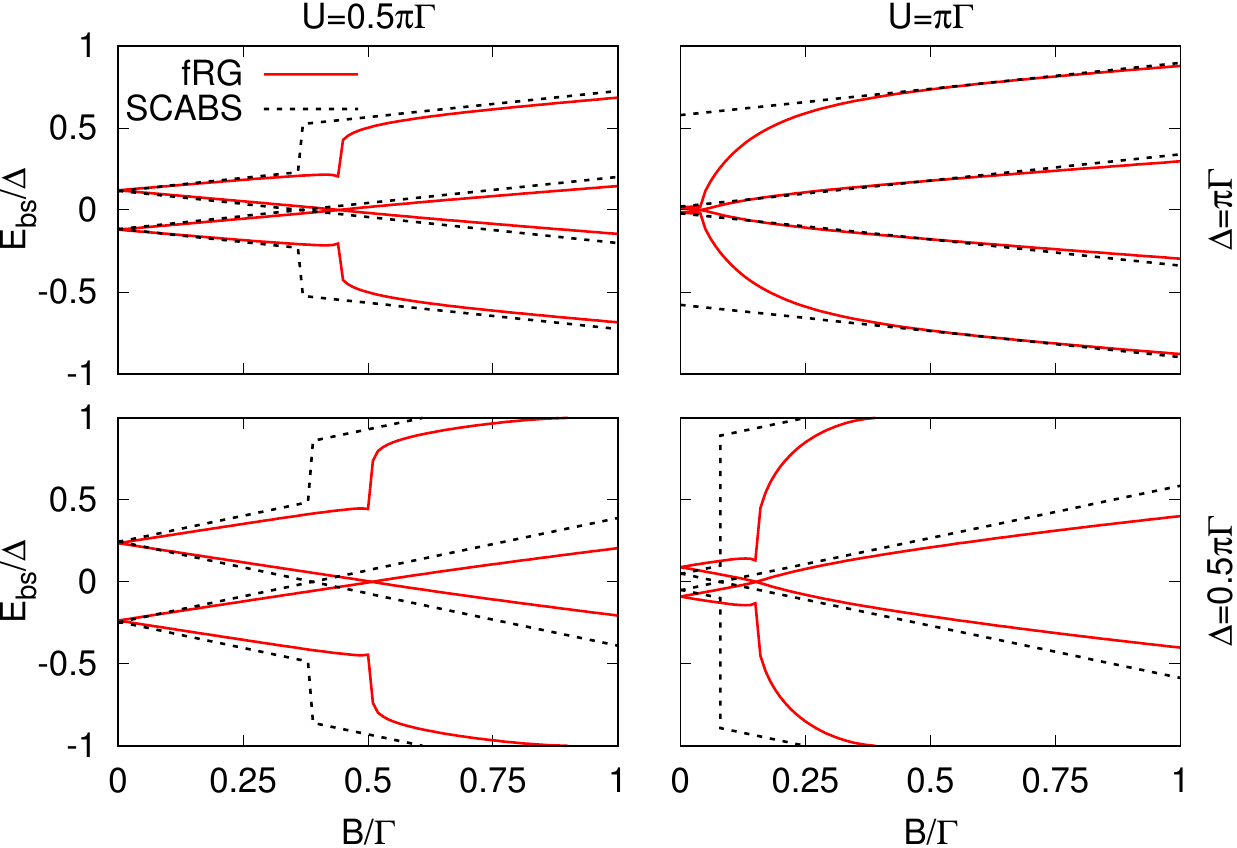}
\caption{ Bound state energies calculated with fRG (full lines) and SCABS approximation
(dashed lines) as a function of $B$ for $\phi=0$, $\epsilon=0$ and different values of $U$ and $\Delta$. }
\label{fig_varB}
\end{center}
\end{figure*}

In this subsection we provide a detailed comparison between the fRG and the
SCABS approximation. While the fRG, being a perturbative approach, is expected to perform
better for smaller values of $U/\Gamma$, the SCABS will by construction perform
better for larger $\Delta/\Gamma$. We have thus chosen $U \in \{ 0.5\pi\Gamma,
\pi\Gamma \}$ and $\Delta \in \{ 0.5\pi\Gamma, \pi\Gamma \}$ for our comparison,
in order to span different ranges of validity of these approaches.
For the other parameters we chose $\epsilon=0$, $\phi=0$ and $B=0$, and then
varied one of these at a time. The corresponding plots can be found in
Figs.~\ref{fig_varEps}-\ref{fig_varB} respectively.

Overall we find a very good quantitative agreement of the results between
the two methods. As expected, the largest deviations can be found for 
$\Delta=0.5\pi\Gamma$ and $U=\pi\Gamma$, since both methods are then pushed
away from their clear regime of applicability.
Varying $\epsilon$, we see an almost perfect agreement for $U=0.5\pi\Gamma$.
Small deviations arise close to the gap edge, which is a trend that continues
throughout the whole comparison. This is tied to a weaker repulsion of the outer
ABS from the gap edge in the SCABS approximation. We also note that for the choice of
parameters $U=\pi\Gamma$ and $\Delta=\pi\Gamma$ we are very close to the
$0$-$\pi$ transition. While the fRG approximation predicts the system to still be in
the 0-phase, SCABS approximation results are already in the $\pi$-phase. 
This tendency of the SCABS approximation towards the $\pi$-phase is also observed 
throughout the whole comparison.

The data with varying $\phi$ shows an artifact of the static fRG calculations
that arises in the absence of a magnetic field. The ABS in the $\pi$-phase for
$B=0$ are not described correctly, but remain pinned at the chemical potential
as they cross the chemical potential at the phase transition, in disagreement
with the SCABS and the previous findings in the atomic limit. This can most
likely be attributed to the static approximation, as in the large-gap
limit the exact self-energy is found to be frequency dependent at zero field 
in the $\pi$-phase. Otherwise the previously described trends hold, and a good
quantitative agreement is achieved in the 0-phase.

As \cfg{fig_varB} shows, increasing the magnetic field $B$ induces the
$\pi$-phase rather quickly, as could already be inferred from the large-gap
phase boundary defined by \ceq{PTCond}. The tendency of the SCABS approximation towards
the $\pi$-phase is clearly visible in the $B$-dependent data, while the fRG
shows a bending of the outer bound states in the $\pi$-phase close to the phase
transition. This latter behavior was also observed in Sec. \ref{sec:LargeGap}
in the comparison to the exact large-gap expressions, and was there identified
as the main deviation. This effect is dominant for small values of the magnetic
field $B$, where the renormalized interaction was found to diverge. In this
limit the truncation of the hierarchy is no longer justified, as it corresponds
to an expansion in the effective interaction. Similar problems using the static
fRG have been found in Ref.~\onlinecite{Karrasch2007}, as the investigated
two-level quantum dot setup was close to degeneracy.

\subsection{Phase diagram at finite $B$}

\begin{figure}[t]
\begin{center}
\includegraphics[width=0.8\columnwidth]{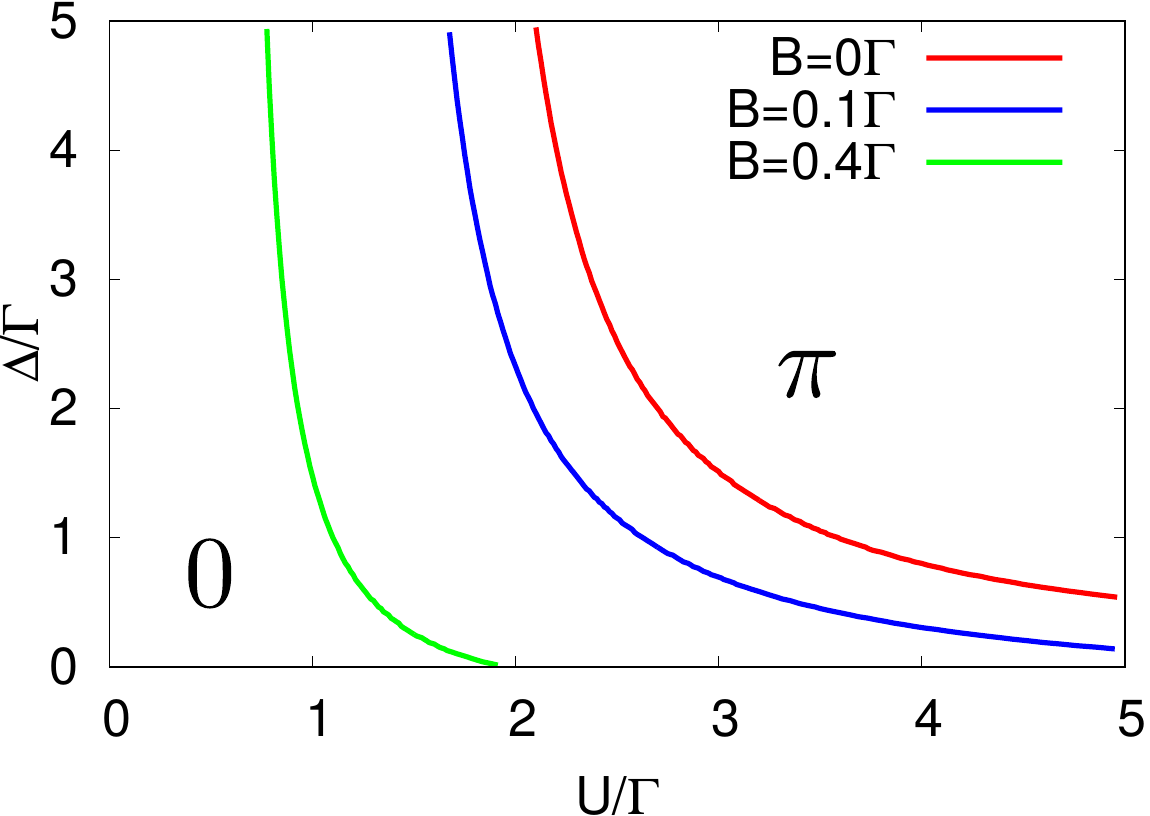}
\caption{Phase diagram as a function of $U$ and $\Delta$ as obtained from the
fRG at $\epsilon=0$ and $\phi=\pi/2$ for different values of $B$. The lines
separate the $0$-phase on the left side from the $\pi$-phase on the right.}
\label{phasediag}
\end{center}
\end{figure}

A detailed phase diagram for the 0-$\pi$ transition determined with fRG is shown in
Fig.~\ref{phasediag}, as a function of Coulomb interaction, gap amplitude,
and several values of the magnetic field (for a choice of phase difference
$\phi=\pi/2$). The general expected trend is a stabilization of the $\pi$-phase 
for increasing values of $U$ and $B$, which both lead to local moment
formation. The $\pi$ state is also favored for increasing values of $\Delta$,
as this removes the quasiparticles and thus weakens the Kondo effect responsible 
for the possible presence of the 0-phase at large $U$.

In experimental setups the magnetic field can be expected to extend beyond the
quantum dot. This effect can lead to a reduction of the superconducting gap in
the leads, which would stabilize the $0$-phase. 

\subsection{Comparison with NRG}
\cfg{compNRG} shows a comparison of fRG data (solid lines) and NRG
data~\cite{Bauer2007} (symbols) for the ABS and the corresponding weights for
$\epsilon = 0$, $B=0$, $\phi=0$ and $\Delta/\Gamma=0.0157, 0.157, 0.9425$ (red,
green, blue). We find a good quantitative agreement with the NRG data up to
interaction values of $U=\pi\Gamma$. For larger $U$ values, frequency
dependent self-energy effects become prominent~\cite{Meng2009}, so that the
static fRG cannot be expected to be precise.

\begin{figure}[b]
\begin{center}
\includegraphics[width=0.8\columnwidth]{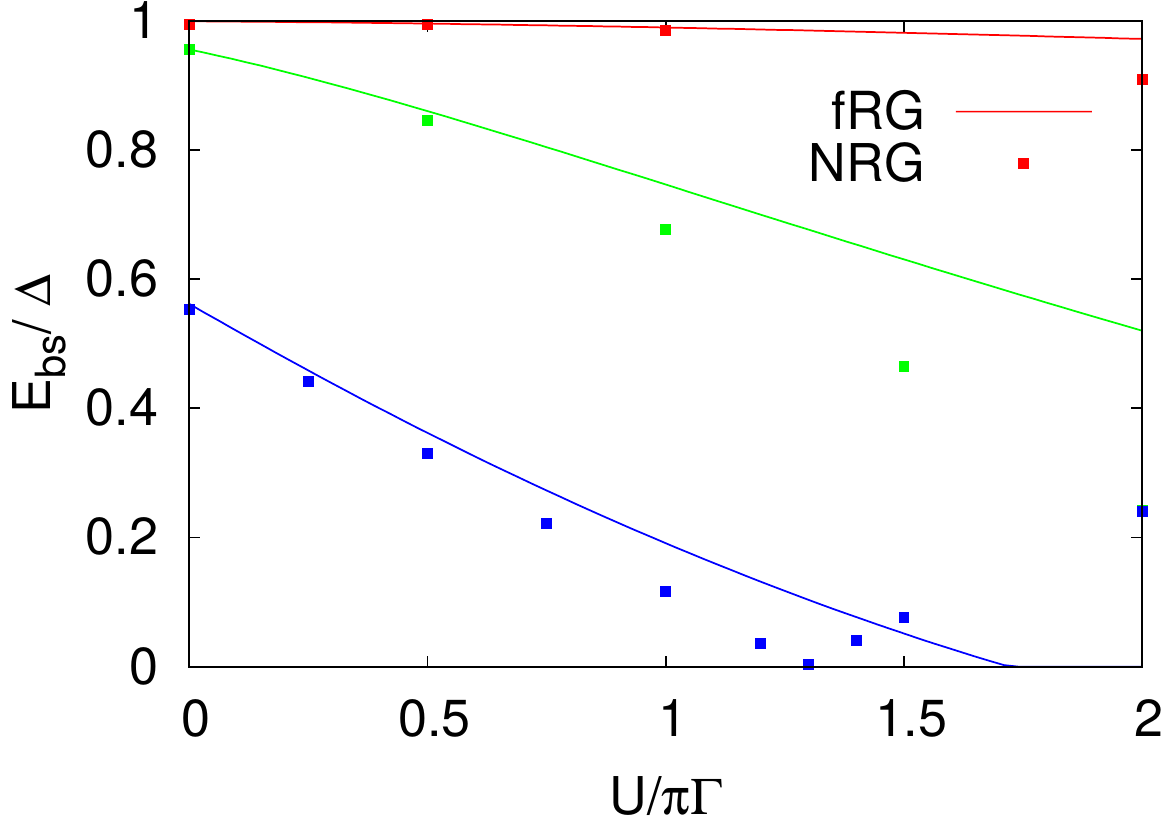}
\hspace{0.00\textwidth}
\includegraphics[width=0.8\columnwidth]{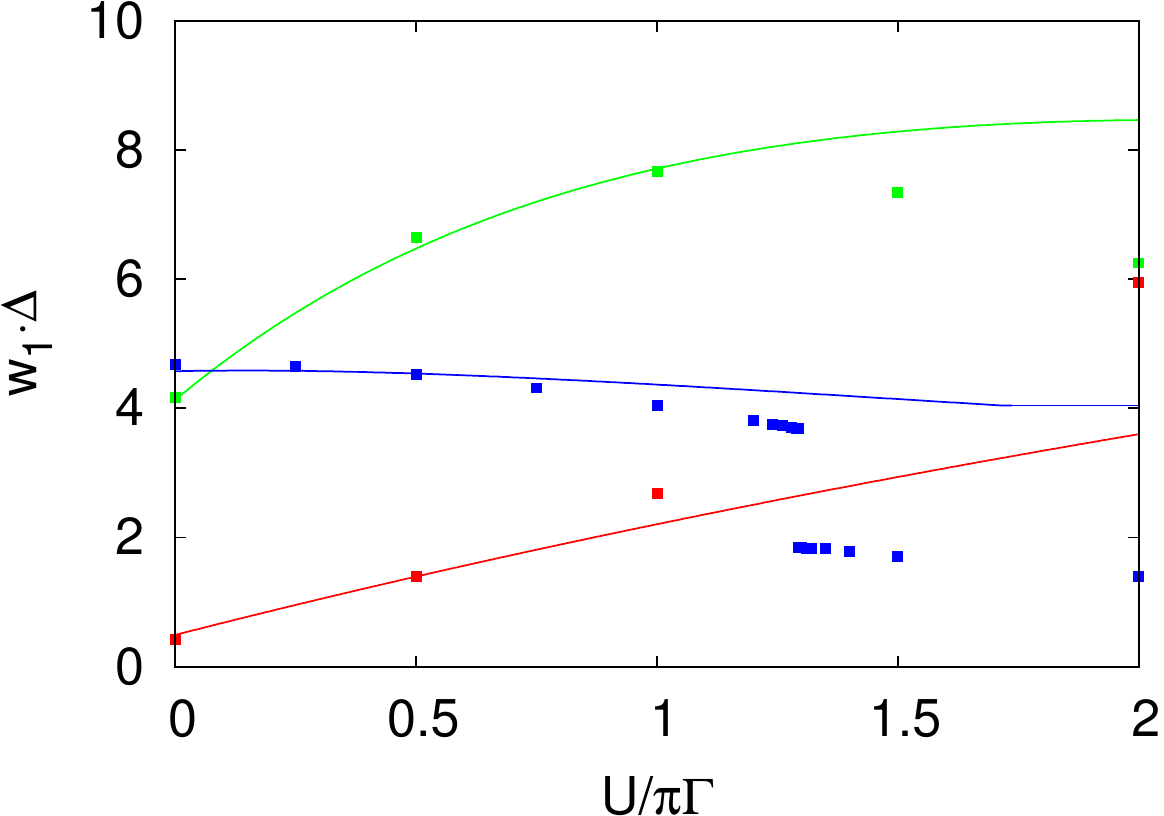}
\caption{Comparison with NRG data from Ref.~\onlinecite{Bauer2007} for the bound
state energy and the corresponding weights as a function of the interaction
strength. The parameters are $\epsilon = 0$, $B=0$, $\phi=0$ and
$\Delta/\Gamma=0.0157, 0.157, 0.9425$ (red, green, blue).}
\label{compNRG}
\end{center}
\end{figure}

\subsection{Spectroscopy}
The density of states in experimental setups like the ones reported in
Refs.~\onlinecite{Pillet2010,Chang2013,Lee2014} is probed by measuring the
differential conductance using a weakly coupled normal lead. This has the effect
that the Andreev bound states are broadened by an energy scale $\Gamma_N$, which
is the corresponding hybridization to the normal contact. This effect can be 
easily accounted for during the fRG flow by considering the additional self-energy
\begin{equation}
\Sigma_N(i\omega) = \begin{pmatrix} -i\Gamma_N \;\rm{sign}(\omega) && 0 \\ 0 
&& -i\Gamma_N \;\rm{sign}(\omega) \end{pmatrix}
\end{equation}
in the Dyson-equation, $G^\Lambda = \left[(G_0^{\Lambda})^{-1} - \Sigma^\Lambda
- \Sigma_N \right]^{-1}$. We can then straightforwardly calculate the density of
states using \ceq{DOS}. One such calculation for a varying on-site energy
$\epsilon$ and $\Gamma_N = 0.1\Gamma$, $\Delta=\Gamma$, $U=3.5\Gamma$ and
$B=0.5\Gamma$ is shown in \cfg{spectDOS}. As expected, the bound states acquire
a broadening due to the presence of the normal lead, and the data compares
qualitatively with measurements from Ref. \onlinecite{Lee2014}.  Note that the
outer bound states in the $\pi$-phase close to $\epsilon=0$ have already been
been absorbed into the continuum, as it can be also observed in Fig.~\ref{fig_varB}.
In view of the experimental observation we point out that the fRG can be easily 
extended to multi-level quantum dot systems.

\begin{figure}[t]
\begin{center}
\includegraphics[width=0.85\columnwidth]{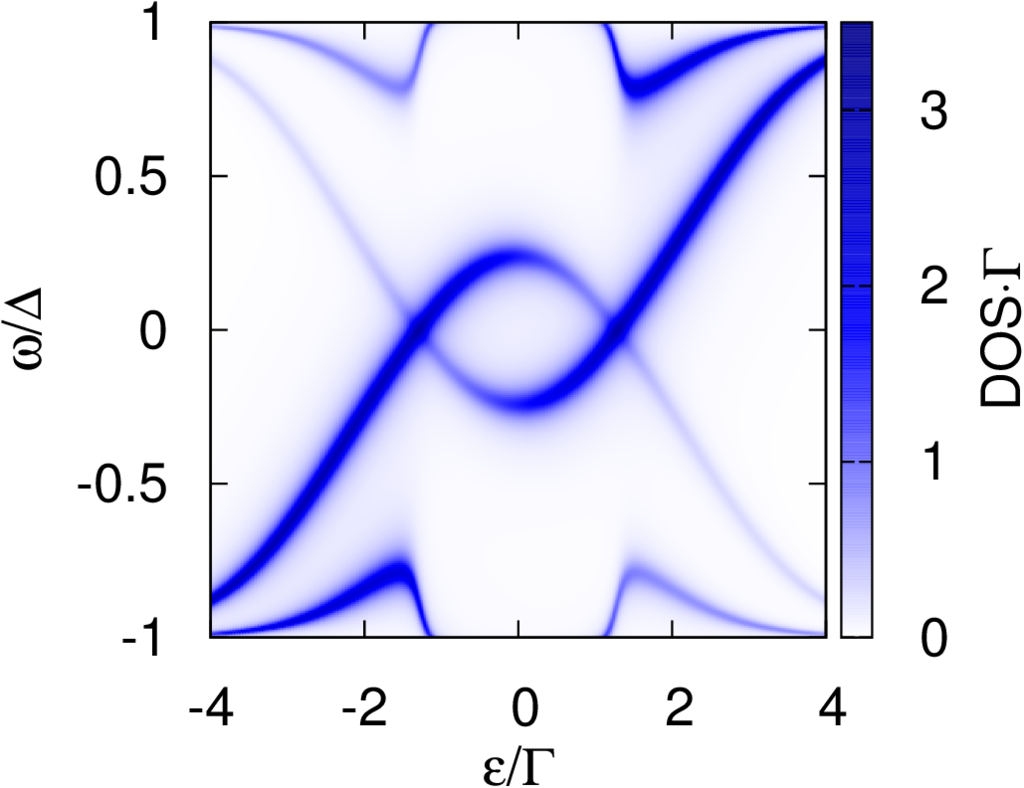}
\caption{fRG results for the density of states as a function of the on-site energy $\epsilon$, for $\Gamma_N = 0.1\Gamma$, $\phi=0$, $\Delta=\Gamma$, $U=3.5\Gamma$, and $B=0.5\Gamma$. The transition from the $\pi$- to the $0$-phase is induced at $\epsilon=\pm 1.5\Gamma$.}
\label{spectDOS}
\end{center}
\end{figure}

\section{Conclusion}

We have investigated electrostatic gating and magnetic field effects on the ABS of 
an interacting quantum dot coupled to superconducting leads by extending the static functional
renormalization group and the self-consistent Andreev bound states theory to
include finite magnetic fields. These complementary approaches
allow to capture the rich physical behavior in the large parameter space with a
reduced numerical effort. According to the range of validity we found a good
quantitative agreement not only between the methods, but also with NRG and the
exact solution in the large-gap limit. The latter was discussed in detail for
the case of a finite magnetic field, allowing for a deeper understanding of
the generic finite-gap situation. We further showed how a local magnetic field induces a
splitting of the ABS whenever the system is $0$-phase, while this effect is
absent in the $\pi$-phase, and provided examples of the tunneling density of states
that is typically measured in experiments.

\acknowledgements
We are grateful to J. Bauer and C. Karrasch for discussions.
We acknowledge financial support from the Deutsche Forschungsgemeinschaft (DFG) 
through FOR 723, RTG 1995, ZUK 63, SFB 1143 and SFB/TRR 21, and the Austrian Science Fund (FWF) within the Project F41 (SFB ViCoM).

\appendix

\section{Green's function in the large gap limit}
\label{app:LargeGap}

To calculate the full Green function in the large-gap limit we use the Lehmann
representation for diagonal correlation functions, which reads
\begin{align}
G_{A A^\dagger} (i\omega) = \sum_{mn} \frac{ |\langle n| A | m \rangle |^2}{E_n - E_m + i\omega} (\rho_n + \rho_m ).
\end{align}
Using the eigenbasis \ceq{ATLIEigen} of the effective Hamiltonian, we find
\begin{align}
G_{b_+ b_+^\dagger} (i\omega) = \frac{\rho_{00} + \rho_{01}}{i\omega-E_{\phi}-B-\frac{U}{2}} + \frac{\rho_{10} + \rho_{11}}{i\omega-E_{\phi}-B+\frac{U}{2}} ,
\label{Gb1}
\end{align}
and
\begin{align}
G_{b_- b_-^\dagger} (i\omega) = \frac{\rho_{00} + \rho_{10}}{i\omega+E_{\phi}-B-\frac{U}{2}} + \frac{\rho_{11} + \rho_{01}}{i\omega+E_{\phi}-B+\frac{U}{2}}.
\label{Gb2}
\end{align}
The off-diagonal elements evaluate to $G_{b_+ b_-^\dagger} = G_{b_- b_+^\dagger}
= 0$.  We now aim at calculating the exact self-energy expressions. For
$B\neq0$, the ground state energy is either $E_{00}$ or $E_{10}$, resulting in
\begin{equation}
\begin{split}
G_{bb^\dagger}^{-1}(i\omega) = i\omega &- \begin{pmatrix} B + E_{\phi}  & 0 \\0 &  B-E_{\phi}  \end{pmatrix} \\
&- \begin{pmatrix} \mp \frac{U}{2} & 0 \\ 0 & \frac{U}{2} \end{pmatrix}, \qquad E_{00} \gtrless E_{10}.
\end{split}
\end{equation}
Using the Dyson equation $G^{-1} = i\omega - H^0 - \Sigma$, we hence obtain
\begin{align}
\Sigma_{b b^\dagger} = \begin{pmatrix} \mp \frac{U}{2} & 0 \\ 0 & \frac{U}{2} \end{pmatrix} , \qquad E_{00} \gtrless E_{10},
\end{align}
for the self-energy. For $B=0$, the 0-phase calculation results in the same
self-energy. For the $\pi$-phase we get
\begin{equation}
\begin{split}
&G_{bb^\dagger}^{-1}(i\omega)\\ 
&= 2 \begin{pmatrix} \frac{1}{i\omega - E_{\phi} - \frac{U}{2}} + \frac{1}{i\omega - E_{\phi} + \frac{U}{2}} & 0 
\\ 0 & \frac{1}{i\omega + E_{\phi} - \frac{U}{2}} + \frac{1}{i\omega + E_{\phi} + \frac{U}{2}} \end{pmatrix}^{-1}\\
&= i \omega - \begin{pmatrix} E_{\phi} & 0 \\0 & -E_{\phi} \end{pmatrix} - \frac{U^2}{4} \begin{pmatrix} \frac{1}{i\omega - E_{\phi}} & 0 \\0 & \frac{1}{i\omega + E_{\phi}} \end{pmatrix}.
\end{split}
\end{equation}
The resulting self-energy
\begin{align}
\Sigma_{b b^\dagger}(i\omega) = \frac{U^2}{4} \begin{pmatrix} \frac{1}{i\omega - E_{\phi}} & 0 \\0 & \frac{1}{i\omega + E_{\phi}} \end{pmatrix},
\label{Sig_B0}
\end{align}
is solely quadratic in the interaction $U$. The corresponding expressions for self-energy and Green functions in the Nambu basis can now be easily acquired by rotating back to the 
old basis. Executing this for the self-energy results in Eqs.~(\ref{SigFinB}) and (\ref{SigBZero}).
The Green function in the Nambu basis can be calculated straightforwardly by the Dyson equation. It will prove more useful though to write
\begin{equation}
\begin{split}
G_{\varphi \varphi^\dagger}&= \begin{pmatrix} u & -v \\ v^* & u^* \end{pmatrix} G_{bb^\dagger} \begin{pmatrix} u^* & v \\ -v^* & u \end{pmatrix}\\
&= G_{b_+ b_+^{\dagger}} \begin{pmatrix} |u|^2 & -u^*v \\ -uv^* & |v|^2 \end{pmatrix} + G_{b_- b_-^\dagger} \begin{pmatrix} |v|^2 & u^*v \\ uv^* & |u|^2 \end{pmatrix},
\label{Gbd}
\end{split}
\end{equation}
since in this representation we can easily read off the bound state weights.

\section{Derivation of the SCABS equations}
\label{app:SCABS}

We here want to summarize, in accordance with Ref.~\onlinecite{Meng2009a}, the derivation of the SCABS equations presented in Sec.~\ref{subsec:SCABS}.
%As already mentioned in Sec.~\ref{subsec:SCABS}, 
Let us begin by considering the hybridization function of the leads for the case of a finite bandwidth $2D$
\begin{equation}
\Gamma_{\phi}(i\omega) = \frac{2}{\pi} \arctan\left(\frac{D}{\sqrt{\Delta^2-(i\omega)^2}}\right) \sum_\alpha \Gamma_\alpha e^{i\phi_\alpha}.
\end{equation}
The non-interacting Green function of the dot then generalizes to
\begin{align}
G_0(i\omega) &= \begin{pmatrix} i\tilde{\omega} - \epsilon - B& \tilde{\Delta}
\\ \tilde{\Delta}^* & i\tilde{\omega} + \epsilon - B\end{pmatrix}^{-1} 
\end{align}
with
\begin{align}
i\tilde{\omega} &= i\omega\left(1+\frac{\Gamma_0(i\omega)}{\sqrt{\omega^2 + \Delta^2}}
\right ), \\
\tilde{\Delta} & = \frac{\Delta}{\sqrt{\omega^2+\Delta^2}} \Gamma_\phi(i\omega).
\end{align}
The system is then fully described by the action
\begin{align}
S &= S_0 + S_{\rm int}
\label{eq:action}
\end{align}
with
\begin{align}
 S_0 &= - \frac{1}{2\pi}\int\operatorname{d\omega}\hspace{0.1cm} \bar{\Psi}(i\omega)^{} G_0(i\omega)^{-1}\Psi(i\omega)^{}.
\end{align}
and
\begin{align}
S_{\rm int} &= - \frac{U}{2\pi}\int\operatorname{d\omega_i}\hspace{0.1cm} \left( \bar{\Psi}_1(\omega_1) \Psi_1(\omega_2) - \frac{1}{2} \right) \\
\times &\left( \bar{\Psi}_2(\omega_3) \Psi_2(\omega_4) - \frac{1}{2} \right ) \delta\left( \omega_1 - \omega_2 + \omega_3 - \omega_4 \right)
\end{align}
in accordance with \ceq{eq:H_dot}.
Here, $\Psi(i\omega)$ and $\bar{\Psi}(i\omega)$ denote the frequency dependent Grassmann-fields corresponding to the previously introduced Nambu-spinors.

We can now decompose the action into a effective part, corresponding to the
limit $\Delta\to\infty$, and all other terms (compare Ref.~\onlinecite{Meng2009a})
\begin{align}
 S = S_{\rm eff} + S_{\rm pert},
\end{align}
with
\begin{subequations}
\begin{align}
 &S_{\rm eff} = - \frac{1}{2\pi}\int\operatorname{d\omega}\hspace{0.1cm} \bar{\Psi}(i\omega)^{} G^{\rm eff}_0(i\omega)^{-1}\Psi(i\omega)^{} + S_{\rm int}, \\
 &G_0^{\rm eff}(i\omega) = \lim_{\Delta\rightarrow\infty} G_0(i\omega),
\end{align}
\end{subequations}
as well as
\begin{align}
 S_{\rm pert} = - \frac{1}{2\pi}\int\operatorname{d\omega}\hspace{0.1cm} \bar{\Psi}(i\omega)^{} \Bigl( G_0(i\omega)^{-1} - G^{\rm eff}_0(i\omega)^{-1} \Bigr) \Psi(i\omega).
\end{align}
Expanding to lowest order in $S_{\rm pert}$ allows to compute straightforwardly the corrections to the atomic levels\cite{Meng2009a}. Note that this formulation in principle
also allows to set up a functional renormalization group flow starting from the exact atomic limit solution, following the ideas of Ref.~\onlinecite{Wentzell2015}.

\bibliography{refs}

\end{document}